\documentclass{aa}

\pdfoutput=1

\usepackage{natbib}
\usepackage{graphicx}
\usepackage{txfonts}
\usepackage{subcaption}
\usepackage{hyperref}
\usepackage{soul}
\usepackage{xcolor}
\usepackage{comment}

\hypersetup{
   colorlinks=true,
   citecolor=blue,
   linkcolor=blue,
   urlcolor=black
   }

\titlerunning{Storms and convection on Uranus and Neptune}
\authorrunning{Clement et al.}

\begin{document}

    \title{Storms and convection on Uranus and Neptune: \\impact of methane abundance revealed by a 3D cloud-resolving model}

    \author{
    Noé Clément \inst{1}
    \and Jérémy Leconte \inst{1}
    \and Aymeric Spiga \inst{2}
    \and Sandrine  Guerlet \inst{2,3}
    \and Franck Selsis \inst{1}
    \and Gwenaël Milcareck \inst{2,4}
    \and Lucas Teinturier \inst{2,3}
    \and Thibault Cavalié \inst{1,3}
    \and Raphaël Moreno \inst{3}
    \and Emmanuel Lellouch \inst{3}
    \and Óscar Carrión-González \inst{3}
    }
    
    \institute{
    Laboratoire d'Astrophysique de Bordeaux, Univ. Bordeaux, CNRS, B18N, allée Geoffroy Saint-Hilaire, 33615 Pessac, France
    \and
    Laboratoire de Météorologie Dynamique (IPSL), Sorbonne Université, Centre National de la Recherche Scientifique, École
    Polytechnique, École Normale Supérieure, Paris, France
    \and
    LESIA, Observatoire de Paris, Université PSL, CNRS, Sorbonne Université, Université de Paris, 5 place Jules Janssen,
    92195 Meudon, France
    \and
    Laboratoire ATmosphère Milieux Observations
    Spatiales/Institut Pierre-Simon Laplace (LATMOS/IPSL), Sorbonne Universités, UPMC Univ Paris 06, Université Paris-Saclay, Université de Versailles
    Saint-Quentin-en-Yvelines, Centre National de la Recherche Scientifique, 78280 Guyancourt, France }

    \date{Submitted December 13, 2023 / Accepted July 26, 2024}

    \abstract
    {
    Uranus and Neptune have atmospheres dominated by molecular hydrogen and helium. In the upper troposphere (between 0.1 and 10 bars), methane is the third main molecule and condenses, yielding a vertical gradient in CH$_4$. This condensable species being heavier than H$_2$ and He, the resulting change in mean molecular weight due to condensation comes as a factor countering convection, traditionally considered as ruled by temperature only. It makes both dry and moist convection more difficult to start. As observations also show latitudinal variations in methane abundance, one can expect different vertical gradients from one latitude to another.
    }
    {
    In this paper, we investigate the impact of this methane vertical gradient and the different shapes it can take, on the atmospheric regimes, especially on the formation and inhibition of moist convective storms in the troposphere of ice giants.
    }
    {
    We develop a 3D cloud-resolving model to simulate convective processes at the required scale. This model is non-hydrostatic and includes the effect of the mean molecular weight variations associated with condensation.
    }
    {Using our simulations, we conclude that typical velocities of dry convection in the deep atmosphere are rather low (of the order of 1 m/s) but sufficient to sustain upward methane transport, and that moist convection at methane condensation level is strongly inhibited.
    Previous studies derived an analytical criterion on the methane vapor amount above which moist convection should be inhibited in saturated environments. In ice giants, this criterion yields a critical methane abundance of 1.2\% at 80~K (this corresponds approximately to the 1~bar level).
    We first validate this analytical criterion numerically.
    We then show that this critical methane abundance governs the inhibition and formation of moist convective storms, and we conclude that the intensity and intermittency of these storms should depend on the methane abundance and saturation.\\
    - In the regions where CH$_4$ exceeds this critical abundance in the deep atmosphere (at the equator and the middle latitudes on Uranus, and all latitudes on Neptune), a stable layer almost entirely saturated with methane develops at the condensation level. In this layer, moist convection is inhibited, ensuring stability. Only weak moist convective events can occur above this layer, where methane abundance becomes lower than the critical value. The inhibition of moist convection prevents strong drying and maintains high relative humidity, which favors the frequency of these events.\\
    - In the regions where CH$_4$ remains below this critical abundance in the deep atmosphere (possibly at the poles on Uranus), there is no such layer. More powerful storms can form, but they are also a bit rarer.
    }
    {In ice giants, dry convection is weak, and moist convection is strongly inhibited. However, when enough methane is transported upwards, through dry convection and turbulent diffusion, sporadic moist convective storms can form.
    These storms should be more frequent on Neptune than on Uranus, because of Neptune's internal heat flow and larger methane abundance.
    Our results can explain the observed sporadicity of clouds in ice giants and can help us guide future observations to test the conclusions of this work.
    }
    
    \maketitle


\section{Introduction}

Uranus and Neptune are the two most distant planets of our Solar System, and thus receive little insolation (3.7 W m$^{-2}$ for Uranus and 1.5 W m$^{-2}$ for Neptune, against 1366 W m$^{-2}$ for the Earth), in addition to having long radiative timescales (more than  100 terrestrial years at 1 bar).
As a result, weak atmospheric activity might be expected, yet observations highlight intense meteorology showing numerous discrete cloud features (presumably composed of methane ice crystals) evolving on short timescales \citep{Karkoschka2011} and long-lasting powerful storms \citep{Molter2018}.
\citet{Hueso2020} listed several observations as candidate moist-convection features in Uranus and Neptune, but they conclude that most cloud activity observed so far is probably not convective. The record of observations demonstrating moist convective activity in Uranus and Neptune is almost nonexistent because frequent observations at very high spatial resolution -- that are lacking today -- would be required.
These clouds and storms need to be modeled in order to understand the atmospheric dynamics of ice giants. A particularly crucial open question is related to the mechanisms of activation and inhibition of convection in those storms.
\par
\begin{figure}[ht!]
    \centering
    \includegraphics[width=\linewidth]{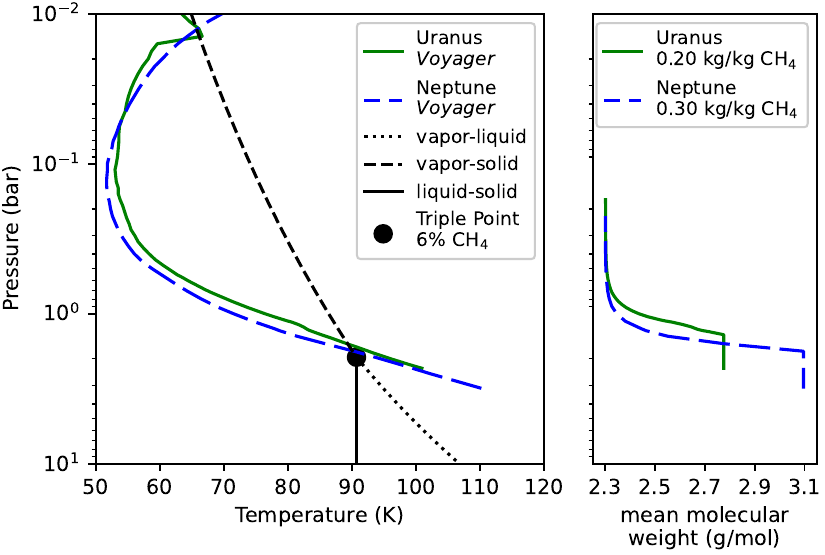}
    \caption{Temperature profiles and variability of mean molecular weight in ice giants. \\
    {\footnotesize
    Left panel: Uranus and Neptune temperature profiles from the \textit{Voyager 2} spacecraft, retrieved from radio occultations by \citet{Lindal1987,Lindal1990,Lindal1992}. 
    Saturation vapor pressure and triple point are indicated for 6\% of CH$_4$ (the maximum observed value), which corresponds to a specific concentration of 0.30~kg/kg.
    Theoretically, methane only exists in a solid or gas form, and can only sublimate and condense. A lower methane concentration would shift the gas/liquid/solid transition curves toward the bottom of the plot.
    \\
    Right panel: Mean molecular weight variability in ice giants. We consider a mix of [85\% H$_2$ + 15\% He] for the background atmosphere composition corresponding to a 2.3~g~mol$^{-1}$ mean molecular weight. We assume that CH$_4$ specific concentration follows saturation and its specific concentration in the deep atmosphere is set to 0.20~kg/kg for Uranus (equivalent to a 3.6\% volume mixing ratio) and 0.30~kg/kg for Neptune (6.2\% volume mixing ratio).}
    }
    \label{fig:U_N_temp_weight}
\end{figure}

Among the properties they share, Uranus and Neptune both have atmospheres dominated by molecular hydrogen and helium, where all condensable species (CH$_4$, H$_2$S, NH$_3$, H$_2$O) are heavier than the background mixture of H$_2$ and He.
Observations show a high abundance of methane in the troposphere, with significant latitudinal variations: 1 to 4\% in Uranus at 2 bars \citep{Sromovsky2014, Sromovsky2018}, 2 to 6\% in Neptune at 4 bars \citep{Irwin2021, Tollefson2019}.
In ice giants, where the troposphere is located below the 0.1~bar level, methane is expected to condense around 1-2 bars (Figure \ref{fig:U_N_temp_weight}). The methane mixing ratio decreases along with pressure between 1-2 bars and  0.1 bar. As a consequence, the mean molecular weight in the atmosphere also decreases along with pressure in these layers (Figure~\ref{fig:U_N_temp_weight}). 
In this study, we explore the 0.03-10~bar pressure range.
Because the other minor species condense much deeper in the troposphere (e.g., H$_2$O around 100 bars) or are much less abundant (e.g., H$_2$S, above the 10 bars level, \citet{Moses2020}), we will consider only methane among the condensable species. 
\par
An abundant and heavy condensable species like methane is thought to play an important role in convective storms that occur in the troposphere.
In the traditional view of convection, temperature alone sets the rules. Relatively to the adiabatic gradient, warmer air rises, and colder air sinks.
By contrast, in ice giants, both temperature and mean molecular weight control convection.
Air at depth is warmer yet heavier; air higher up is colder yet lighter. This makes convection in ice giants more complex than in the traditional view.
\citet{Guillot2019} showed that the potential temperature
increase required to compensate for the mean molecular weight change and have a density profile that is neutrally stable to convection is 20~K, not including possible latent heat effects.
During flybys of the ice giants, the \textit{Voyager 2} spacecraft has measured vertical temperature profiles (Figure \ref{fig:U_N_temp_weight}), showing near-dry-adiabatic gradients in the troposphere for both Uranus and Neptune \citep{Lindal1987,Lindal1990,Lindal1992}.
Since the thermal gradient is close to the dry-adiabatic gradient,
it is relevant to address the impact of the mean molecular weight vertical gradient. It can completely change the (un)stability of the atmosphere.

Global Climate Models (GCM) have been used to model the atmosphere of Jupiter \citep{Boissinot2024} and Saturn \citep{Guerlet2014, Li2015, Spiga2020, Bardet2020}, highlighting large-scale phenomena but also the need to include mesoscale processes that GCM cannot directly resolve.
Indeed, GCMs for ice giants have a horizontal resolution of 1° in latitude at best (equivalent to 400~Km), in addition to being vertically hydrostatic.
Because of this assumption of hydrostaticity, a GCM cannot be used to study convective storms.
Solving and studying convection requires models with a higher spatial resolution and capable of solving the vertical momentum equation without making the hydrostatic approximation.
\par
\citet{Sugiyama2014} and \citet{Li2019} have studied moist-convection on Jupiter with cloud-resolving models. In Jupiter, the condensable species (H$_2$O, NH$_3$, H$_2$S) also have a molecular weight higher than H$_2$ and He.
\citet{Sugiyama2014} conclude that: \\
\noindent - stable layers associated with condensation act as effective dynamic boundaries, \\
\noindent - intense cumulonimbus clouds develop with clear temporal intermittence, with a period that is roughly proportional to the deep abundance of H$_2$O gas,\\
\noindent - the active transport associated with these clouds leads to the establishment of mean vertical profiles of condensates and condensable gases that are markedly different from the usual three-layer structure. \\
However, in none of their simulations condensable species (H$_2$O, NH$_3$, H$_2$S) exceed their critical specific concentrations (see next section) for moist convection inhibition.
The simulations of an idealized Jovian atmosphere in radiative-convective equilibrium presented by \citet{Li2019} show that the temperature gradient is super-adiabatic near the water condensation level because of the change of mean molecular weight. \\
Using non-hydrostatic simulations applied to Uranus and Neptune, \citet{Ge2024} show that CH$_4$ and H$_2$S condensation induces two stably stratified layers at about 1 bar and 8 bars when the abundance of these elements range from 30 times solar to 50 times solar. They find that, in these stable layers, the temperature profile is super-adiabatic and convection is inhibited, because of the compositional gradient in sub-saturated weather layers.
More generally, they find that weakly forced giant planets are less cloudy than previously expected and that moist convection is limited by the planetary heat flux.

\par
To study how the mean molecular weight variability affects convection, the challenge is to build a model that can resolve convection and account for the effect of condensation on the mean molecular weight.
In this paper, using a 3D cloud-resolving model, we investigate the impact of vertical mean molecular weight gradients, induced in particular by methane condensation, on the formation and inhibition regimes of convective storms in ice giants. 
\par
In Section \ref{citeria}, we review the criteria for convection inhibition in ice giants. In Section \ref{model} we present the model, how we adapted it for this study, and which challenges it implies. In Section \ref{dynamics}, we describe the simulated atmospheric structures, and we analyze their temporal evolution in Section \ref{intermittency}. In Section \ref{discussion}, we discuss the limitations of our model and several open questions about the understanding of ice giants, and we attempt to give an overall scenario for storm formation based on the results of our simulations.


\section{Convection inhibition criteria in ice giants} \label{citeria}

\begin{figure*}[ht!]
\centering
    \includegraphics[width=0.8\linewidth]{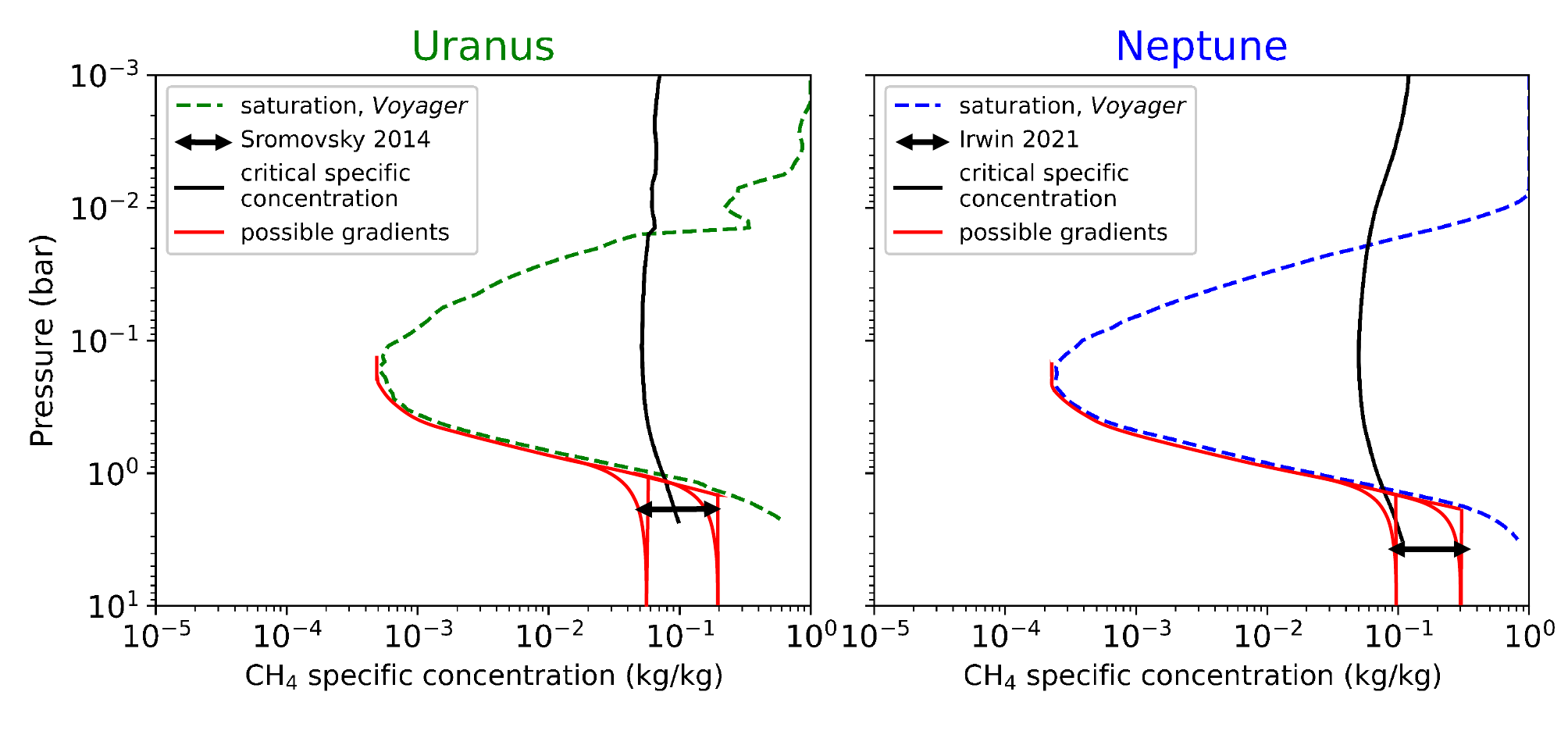}
    \caption{Possible methane vapor profiles in ice giants. \\
    In red we plot several possible methane vapor profiles constrained by observations from \citet{Sromovsky2014} and \citet{Irwin2021} (we plot the range of the latitudinal variations) and by saturation curves corresponding to \textit{Voyager 2} temperature profiles. \\
    Criterion curves in black indicate when moist convection is inhibited: when the specific concentration is higher than the critical specific concentration and the level is saturated, then the criterion applies.
    These possible methane vapor profiles are inspired by the tested profiles in \citet{Sromovsky2018} who talk about "descended profiles".
    }
    \label{fig:U_N_methane}
\end{figure*}

\citet{Guillot1995} and \citet{ Leconte2017} calculated analytical criteria for convection inhibition caused by a vertical gradient of the mean molecular weight.
To determine if convection can occur for a given thermal gradient, the density of a theoretical rising parcel should be compared to its surrounding environment. The rising parcel follows the adiabatic gradient, dry if the pressure level is not saturated, and moist if it is. In a dry and well-mixed environment, the parcel continues to rise (i.e. convection occurs) if the thermal gradient is higher in absolute value than the dry-adiabatic gradient. The parcel remains warmer than the environment, with lower density and positive buoyancy.
In the presence of a gradient of methane (i.e. there is more methane at the bottom than at the top), convection can be inhibited even if the thermal gradient is super-adiabatic.

\par At dry levels, if there is a gradient of methane, a rising parcel coming from the bottom, contains more methane than the air at the top. The density of the parcel is increased relative to the density of the environment. If the methane gradient is strong enough, the density of the parcel becomes higher than the density of the environment, even if the parcel is warmer than the environment. The buoyancy of the rising parcel is negative, stopping convection.
The criterion for dry convection inhibition is the Ledoux criterion: 

\begin{equation}
    \nabla_T < \nabla_{\text{ad}} + \nabla_{\mu}
\end{equation}

\noindent
where $\nabla_T=\frac{\text{d} \ln T}{\text{d} \ln P}$ is the thermal gradient of the atmosphere, $\nabla_{\text{ad}}=\left.\frac{\partial \ln T}{\partial \ln P}\right|_{\text{ad}}$ the dry-adiabatic gradient and $\nabla_{\mu}=\frac{\text{d} \ln \mu}{\text{d} \ln P}$ the mean molecular weight gradient of the atmosphere, ($T$,$P$) being the temperature versus pressure profile of the atmosphere.
\\
\noindent
Here the mean molecular weight gradient and the thermal gradient come as additive factors: the greater the mean molecular weight gradient, the greater the thermal gradient required to trigger convection.

\par In the "moist" atmosphere, methane condenses, producing a vertical methane gradient.
The resulting methane profile is close to the saturation vapor curve. A rising parcel follows the moist-adiabatic gradient: it is becoming cooler, with a decreasing methane abundance because of condensation.
When the mean molecular weight is vertically constant (i.e. in an atmosphere where the condensable species would have the same weight as the background atmosphere), if the thermal gradient is steeper than the moist-adiabatic gradient, then the parcel continues to rise (i.e. convection occurs). (Warming of the rising parcel by latent heat release is included in the moist-adiabatic gradient, which is lower than the dry-adiabatic gradient.)
\\
\noindent
When the mean molecular weight decreases with pressure, if the thermal gradient is steeper than the moist-adiabatic gradient, the abundance of methane also decreases with pressure faster in the environment than in the parcel.
\\
\noindent
A super-moist-adiabatic thermal gradient leads to an inner competition. On the one hand, the steep temperature gradient and the release of latent heat favor convection, and on the other hand varying mean molecular weight prevents convection. The rising parcel is warmer but also heavier than the surrounding environment. The abundance of methane determines which factor dominates, and convection is inhibited if the methane abundance exceeds a critical specific concentration $q_{\text{cri}}(T)$. \citet{Leconte2017} provide the criterion. Moist convection is inhibited if: 

\begin{equation}
(\nabla_T - \nabla_{\text{ad}}^{*})(q_{\text{v}} - q_{\text{cri}}(T))>0
\end{equation}

\noindent
where $\nabla_T$ is the thermal gradient of the atmosphere,
$\nabla_{\text{ad}}^*$ is the moist-adiabatic gradient,
$q_{\text{v}}$ (kg/kg) is the specific concentration of vapor,
and $q_{\text{cri}}(T)$ is the critical specific concentration:
\begin{equation}
    q_{\text{cri}}(T) = \frac{1}{1-\frac{M_\text{gas}}{M_{\text{cs}}}}\frac{R}{\Delta H_{\text{cs}}}
     T = 0.078 \hspace{0.1cm} \frac{T}{80~K} \text{(kg/kg)}
\end{equation}

\noindent
where $M_\text{gas}=2.3$ g mol$^{-1}$ is the mean molecular weight of the non-condensing atmosphere (here a mix of 85\% of H$_2$ and 15\% of He),
$M_{\text{cs}}=16.04$ g mol$^{-1}$ the molecular weight of the condensable species (methane in this study),
$\Delta H_{\text{cs}}=10000$ J mol$^{-1}$ is the latent heat of sublimation (or vaporization depending on the pressure-temperature range) of the condensable species, methane in this study,
$R=8.314$ J K$^{-1}$ mol$^{-1}$ is the perfect gas constant.

\noindent
If $q_{\text{v}}$ exceeds $q_{\text{cri}}(T)$, moist convection is inhibited, even if the thermal gradient is stronger than the moist-adiabatic gradient. 
In this case, the vapor abundance does not come as an additional factor, as it was through the gradient $\nabla_\mu$ in the criterion for inhibition of dry convection, but as a multiplicative factor $(q_{\text{v}} - q_{\text{cri}}(T))$.
Contrary to dry convection inhibition, no thermal gradient however strong can produce moist convection.
This critical specific concentration is linearly dependent on temperature. As the condensation level is around 80~K in ice giants, we can keep in mind the value of 0.078~kg/kg. This corresponds to a volume mixing ratio of 1.2\%.
In \citet{Leconte2017} the value of 0.10~kg/kg at 80~K was proposed. This value was calculated with the latent heat of vaporization (instead of sublimation). As shown by Figure \ref{fig:U_N_temp_weight}, considering solid-gas equilibrium (i.e. using the latent heat of sublimation in the formula) is more adequate.
In Figure \ref{fig:U_N_methane}, we plot several possible methane vapor gradients constrained by observations and saturation curves.
We can see that the criterion of moist convection inhibition should apply at some depth and latitude on both planets.


\section{A 3D cloud-resolving model} \label{model}

The Generic Planetary Climate Model (Generic PCM) gathers different versions of a common structure to model the atmospheres of planets and moons of our Solar System \citep{Spiga2009,Lefevre2017, Bardet2020, Boissinot2024}, as well as exoplanets \citep{Turbet2022}. PCM is the new name of the planetary versions of the model known so far as the Laboratoire de Météorologie Dynamique (LMD) model. Our model is one of these versions.

To build it, two components are coupled:
\begin{itemize}
    \item a dynamical core, which solves the Euler equations of motion
    \item a physical package (by physical, we mean everything that is not related to dynamics), which calculates the tendencies of the relevant physical phenomena
\end{itemize}

\subsection{The dynamical core - The Weather Research and Forecasting (WRF) Model}

In this study, the dynamical core used is adapted from the Weather Research and Forecasting (WRF) Model. This model has been developed for a few decades for meteorological applications on Earth (large-eddy simulations) and used by a lot of meteorological research institutes. The version we use is the 4th one \citep{Skamarock2019}.
To build our own model, we remove the physical package of the WRF model to keep only the dynamical core, which is coupled to our own physical package.
\par
The dynamical core discretizes and solves the Euler equations of motion in a 3D rectangular grid. In addition to the classic fluid mechanics terms, these equations contain the physical tendencies of the relevant phenomena in the atmosphere, such as radiative transfer, for example, calculated by the physical package.
As it is a cloud-resolving model designed for large-eddy simulations, with this chosen dynamical core, we can solve cloud formation. Clouds (identified here as saturated levels), being bigger than the resolution of the model, are spread over several grid cells.
The WRF dynamical core solves dynamics and transport.
It has several characteristics that are particularly interesting in our case:

\begin{itemize}
\item it uses the formalism by \citet{Laprise1992} for the coordinates. The hydrostatic component of dry air pressure is used as the vertical coordinate.
\item it is non-hydrostatic, so it can solve convection.
\item it takes into account, in its equations, the variability of the mean molecular weight by including the variability of air density due to moisture \citep{Leconte2024}.
\end{itemize}

Some studies have already been done on Venus and Mars, using the same dynamical core with the adequate physical package from the PCM for each planet. Simulations on Mars \citep{Spiga2009, Spiga2017} have demonstrated that localized convective snow storms can occur, and simulations on Venus \citep{Lefevre2017,Lefevre2018} have reproduced the vertical position and thickness of the main convective cloud layer. In parallel with this study, the same association of the WRF dynamical core with the physical package we use here has been made for K2-18b by \citet{Leconte2024} who give more details on the dynamical equations.

\subsection{The physical package}

The physical package is a "generic" physical package, which can be used for giant planets \citep{Bardet2020, Guerlet2020}, paleoclimates of telluric planets and exoplanets \citep{Turbet2022}.
Our physical package takes into account the following phenomena:

\begin{itemize}
    \item Radiative transfer (absorption and emission by the gases with the correlated-k formalism; absorption, emission and scattering by aerosols layers; Collision-Induced Absorption (CIA); Rayleigh scattering). Considered species are H$_2$, He, CH$_4$, C$_2$H$_2$, C$_2$H$_6$.
    \item Methane thermodynamic cycle as a condensable species: condensation and sublimation with latent heat release, condensates precipitation.
    \item For Neptune, an internal heat flow (0.43 W m$^{-2}$, \citet{GuillotGautier2015}), which is a residual heating of the gravitational contraction of the planet, and is taken into account as a heat source at the bottom of the model.
\end{itemize}

For the radiative transfer, we use the parameters prescribed for ice giants by \citet{Milcareck2024}, which integrate the aerosol scenario described by \citet{Irwin2022}. Radiative-convective equilibrium 1-D simulations produce temperature profiles close to the observed ones.
Contrary to \citet{Milcareck2024} who used a fixed methane vertical profile, the methane abundance can vary in our simulations (due to condensation, sublimation, and transport). Radiative transfer calculations are updated during the simulation to take into account these methane variations.
\par
The treatment of the methane cycle is generic and can be used for any condensable species, as it has been done to study cloud formation on the Hot-Jupiter WASP-43b by \citet{Teinturier2024}.
This numerical thermodynamic scheme ensures that the methane abundance never exceeds the saturated value, which is derived from saturation pressure calculated with the Clausius-Clayperon formula.
As soon as there is too much vapor at a given pressure-temperature level, the scheme condenses enough methane to bring the vapor amount below the saturation.
The so-formed condensates then precipitate whenever their specific concentration exceeds a threshold that we keep arbitrarily small in this first study, so as not to have to take cloud radiative feedback into account.
The condensates, which are ice crystals, precipitate and are transported instantaneously to deeper unsaturated layers where they sublimate. To do so, at each time step, the routine starts from the level where the condensates have formed and carries them downward. At each unsaturated layer, the mass of methane that is needed to bring the layer back to saturation is computed accounting for the thermal effect of the sublimation. This mass is then sublimated into gas before repeating the process in the next layer until no more precipitations remain.
We constrain the methane profile by a fixed value in the deep atmosphere ($q_\text{deep}$(kg/kg)), that will be chosen among the possible ones allowed by observations. This boundary condition works as an infinite source and sink of CH$_4$ that always maintains its abundance at the bottom of the model at this fixed value.

\subsection{Simulation settings}

\begin{table}[ht!]
\centering
\caption{Settings shared by all simulations}
\begin{tabular}{|l|c|}
     \hline
     \textbf{Parameters} & \textbf{Values} \\
     \hline
     pressure range & 0.03-10 bars  \\
     \hline
     vertical levels & 200  \\
     \hline
     horizontal resolution (dx,dy) & 2 km  \\
     \hline
     horizontal grid points & 50$\times$50  \\
     \hline
     insolation & 0.93 W m$^{-2}$ for Uranus \\
     & 0.38 W m$^{-2}$ for Neptune  \\
     \hline
     internal heat flow & None for Uranus \\
     & 0.43 W m$^{-2}$ for Neptune \\
     \hline
     heat capacity & 10200 J kg$^{-1}$ K$^{-1}$ \\
     \hline
     horizontal boundary conditions & cyclic \\
     \hline
     duration & 1 terrestrial year \\
     & (365$\times$86400s) \\
     \hline
     dynamical time step & 5s \\
     \hline
\end{tabular}
\label{table:ini_conditions}
\end{table}

The parameters common to all simulations are summarized in Table \ref{table:ini_conditions}.
We choose to set the bottom pressure level at 10~bars to have enough pressure levels below methane condensation. We set the top of the model at 0.03~bar to encompass the tropopause, which is just below the 0.1 bar level, and the lower stratosphere.
We have defined 200 vertical pressure levels from 10 bars to 0.03 bar, with an almost regular distribution in log pressure. This pressure distribution allows us to have enough levels where condensation occurs. It corresponds to an average vertical resolution of 0.75~km. The choice of the dynamical time step is constrained by Courant–Friedrichs–Lewy condition. In our case, 5~s is an optimized choice.
With this choice, there is no need to set a vertical-velocity damping in the first 190 levels.
In the top 10 levels, which correspond to the range 0.03-0.05 bar, we set an implicit gravity-wave "damping" layer (\verb|damp_opt = 3|, \verb|dampcoef = 0.2|; these parameters are detailed in the Modeling System User's Guide of WRF whose reference is given in \citet{Skamarock2019}) that smoothes out high speeds to prevent the model from collapsing.
When velocity damping happens, energy is lost. However, as discussed in Section 2.4 of \citet{Leconte2024}, this energy sink is rather small (less than 2\% of the global budget).
In our model, the stratosphere starts at the 0.2 bar level (see Section \ref{dynamics}). Between the 0.2 bar level (the bottom of the stratosphere) and the 0.05 bar level (the bottom of the damping layer), there are 30 vertical grid points. These 30 points are enough for gravity waves to propagate.
Stratospheric dynamics will not be studied here, as this artificial damping reduces and modifies them, and because the chosen pressure range and discretization of the model are not suited to study them. The last layer of the model at the 0.03 bar level is directly subjected to the solar flux, without attenuation by the non-simulated layers above, in order to have a complete radiative balance.
The WRF dynamical core requires a few specific settings.
We have decided to turn off any subgrid parameterization of diffusion (\verb|khdif = 0|, \verb|kvdif = 0|) and hyperviscosity (\verb|c_s = 0|) to ensure that any mixing that we see in the simulations is due to the resolved dynamics of the dynamical core.
The only diffusive processes left in the dynamical core are divergence damping (\verb|smdiv = 0.1|) and external model filters (\verb|emdiv = 0.01|) that are necessary to stabilize the simulations. These are canonical values recommended by the WRF documentation.

\par
\begin{figure}[ht!]
    \centering
    \includegraphics[width=0.8\linewidth]{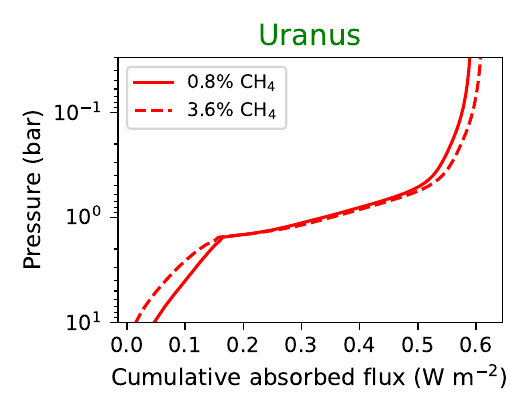}
    \caption{
    Absorbed flux by Uranus' atmosphere.\\
    We test two different methane abundances in the deep atmosphere (0.8\% and 3.6\%).
    We calculate the cumulative sum (from bottom to top) of the flux. The value at the top is the total absorbed flux, the so-called ASR.
    }
    \label{fig:Uranus_absorbed_flux}
\end{figure}

To set the heat capacity, which is constant in our model, we have chosen a value that gives the best fit of a dry-adiabatic temperature profile to \textit{Voyager 2}'s temperature profiles with an associated molecular weight of non-condensing air (85\% H$_2$ + 15\% He) $M_\text{gas}=2.3$ g mol$^{-1}$.
We take $c_p = 10200$ J kg$^{-1}$ K$^{-1}$.
The chosen horizontal resolution is 2~km. It is a good compromise between high resolution and sufficient horizontal coverage, with a 100~km-width domain (50$\times$2~km). Given the horizontal size, the effects of the Coriolis force will not be studied  (they appear at larger scales).
Diurnal and seasonal effects are turned off.
Their impact will be addressed in Section \ref{discussion}.
We run the simulations with the diurnal-averaged insolation on the planet, which corresponds to 1/4 of the solar flux reaching the planet's orbit.
While Uranus has a very low internal heat flow (0.042$^{+0.047}_{-0.042}$ W m$^{-2}$, \citep{Pearl1990,Pearl1991,GuillotGautier2015}), Neptune has a higher one (0.433$\pm$0.046 W m$^{-2}$). We set Uranus' internal heat flow to zero and Neptune's one to 0.43 W m$^{-2}$.
Calculations of radiative transfer in our model show that a bit less than 0.1 W m$^{-2}$ of the incoming solar flux (0.93  W m$^{-2}$) penetrates the atmosphere below the condensation level (see Figure \ref{fig:Uranus_absorbed_flux}).
For a low methane abundance in the deep atmosphere (0.8\%), there is even about 0.05 W m$^{-2}$ absorbed below the model bottom.
These fluxes are enough to trigger convection in the deep atmosphere.
In the model, this energy is assumed to be absorbed at the model bottom and is sufficient to power convection from there.

\subsection{Initialization and convergence}

\begin{table*}[ht!]
\begin{center}
\caption{Varying parameters related to methane between simulations.}
\label{table:ini_qvap_deep}
\begin{tabular}{|l|c|c|l|}
     \hline
     Planet & $q_\text{deep}$ - methane specific concentration & equivalent methane volume mixing ratio & moist convection \\
     &  in the deep atmosphere & in the deep atmosphere & inhibition criterion \\
     \hline
     Uranus & 0.05~kg/kg & 0.8\% & unsatisfied \\
     \hline
     Uranus & 0.20~kg/kg & 3.6\% & satisfied \\
     \hline
     Neptune & 0.05~kg/kg &  0.8\% & unsatisfied \\
     \hline
     Neptune & 0.30~kg/kg & 6.2\% & satisfied \\
     \hline
\end{tabular}
\end{center}
{\footnotesize
The methane specific concentration (kg/kg) is the mass of methane divided by the total mass. The methane volume mixing ratio (\%) is the number of methane molecules divided by the number of the other (non-condensing) molecules (H$_2$ and He).}
\end{table*}

Ice giants are weakly forced systems because of their long radiative time scales and the little insolation they receive.
We therefore expect long convergence times - decades to centuries - for the 3D simulations to reach a steady state.
A simulation of one terrestrial year with our 3D model requires two weeks of computation on a cluster. Simulating a few terrestrial years would consequently extend the duration of computation while remaining a short time compared to the radiative time scales.
We decide to limit ourselves to this duration of one terrestrial year, being aware that our simulations may not reach a statistical steady state. This limitation must be kept in mind when interpreting the results as discussed later in Section \ref{discussion}.
\par
To mitigate this issue, the simulations are started as close as possible to the envisioned equilibrium state. \citet{Leconte2024} have simulated temperate exo-Neptunes for which thermal equilibrium can be reached much faster, because of shorter radiative timescales.
They show that a simple 1D model can well predict the behavior of the thermal profile of 3D simulations. To initialize the 3D simulations with 1D thermal profiles, we have done preliminary work on 1D simulations at radiative-convective equilibrium using their approach.
These 1D simulations use the same single-column physics package as the 3D model, for the radiative and microphysical considerations.
Concerning the dynamics of the 1D simulations, we parameterize convective adjustment and turbulent vertical diffusion.
The convective adjustments are performed by two schemes.
The 1D dry convective adjustment scheme brings back any decreasing profile of virtual potential temperature (this concept will be introduced in the next section) to a constant profile, in dry layers, and mixes the methane accordingly.
The 1D moist convective adjustment scheme is triggered in regions where i) the thermal gradient is steeper than the moist-adiabatic gradient, ii) methane is at saturation, and iii) the methane abundance is lower than the critical abundance discussed in Section \ref{citeria}. In these regions, the thermal gradient is brought back to the moist one and methane is condensed accordingly.

\noindent
The 3D simulations that we will present in the sections hereafter are initialized with 1D simulations that use these parametrizations.
To test the sensitivity to the initial conditions, 3D simulations initialized with 1D simulations that only use a dry convective adjustment have also been run and will be discussed in Section \ref{discussion}.
\par
In order to represent a diverse but limited set of conditions, we have chosen different configurations for initial methane profiles.
In 1D simulations, the parameterizations (convective adjustment and turbulent vertical diffusion) and the constraints due to the saturation vapor curve, build the methane profile. It results in a constant profile at the value set by $q_\text{deep}$, from the bottom of the simulation to the condensation level, and above this level, the profile follows the saturation vapor curve.
In this 1D initialization profile, the methane mixing ratio above the cold trap is constant at its value at the cold trap.
Some observations show a decreasing methane gradient in Uranus' stratosphere and an excess of methane in Neptune's stratosphere \citep{Lellouch2015}. We do not take these variations into account, as their study would require other simulation settings.
\par
The parameter $q_\text{deep}$ is the one that allows us to test different configurations, that are inspired by both observations and analytical criteria.
Observations show a latitudinal minimum around the 2 bar level of about 1\% methane on Uranus and 2\% methane on Neptune. The critical mixing ratio for moist convection inhibition being 1.2\% (at 80~K), we have chosen to include a case study with only 0.8\% methane ($q_\text{deep}$ = 0.05~kg/kg) in order to be below the critical mixing ratio at all pressures. In the case of Neptune, observations show that this configuration might not exist, however, we keep it as an experiment.
We also simulate atmospheres with more methane in the deep atmosphere: 3.6\% for Uranus ($q_\text{deep}$ = 0.20~kg/kg) and 6.2\% for Neptune ($q_\text{deep}$ = 0.30~kg/kg).
Although there might be a (high) pressure level at which CH$_4$ is latitudinally homogeneous, our simulations do not attempt to capture this aspect which will be discussed later.
Finally, we run 4 different simulations in ice giants (2 on Neptune and 2 on Uranus). Table \ref{table:ini_qvap_deep} synthesizes the varying parameters related to methane between these simulations.
\par
After initialization with a 1D profile, we run 3D simulations for one terrestrial year. They need a few simulated months to reach a steady state. From the one terrestrial year simulation, we keep only the last 200 days, from day 150 to day 350, for our study.

\subsection{3D validation tests}

\par
Before using our 3D model in specific configurations, we run more theoretical 3D simulations to check the application of our model to the ice giants.
The first test simulates a dry atmosphere, where we remove the thermodynamic cycle of methane as a condensing species, only keeping it as a radiatively active species. The atmosphere is then only a radiative/dry-convective atmosphere. As expected, this test simulation exhibits a dry convective layer driven by radiative heating at depth overlain by a stratosphere.
The second test simulates a saturated moist atmosphere where we treat methane as a condensing species, but do not account for its different mean molecular weight. Again, as expected, this test simulation exhibits the standard 3-layer structure, with a moist convective layer between the dry troposphere and the stratosphere.

These two tests are conclusive and confirm that the model can be configured to simulate the ice giants.


\section{Structure and dynamics in simulated atmospheres} \label{dynamics}

\begin{figure*}[ht!]
\begin{subfigure}{0.5\textwidth}
  \centering
  \includegraphics[width=\linewidth]{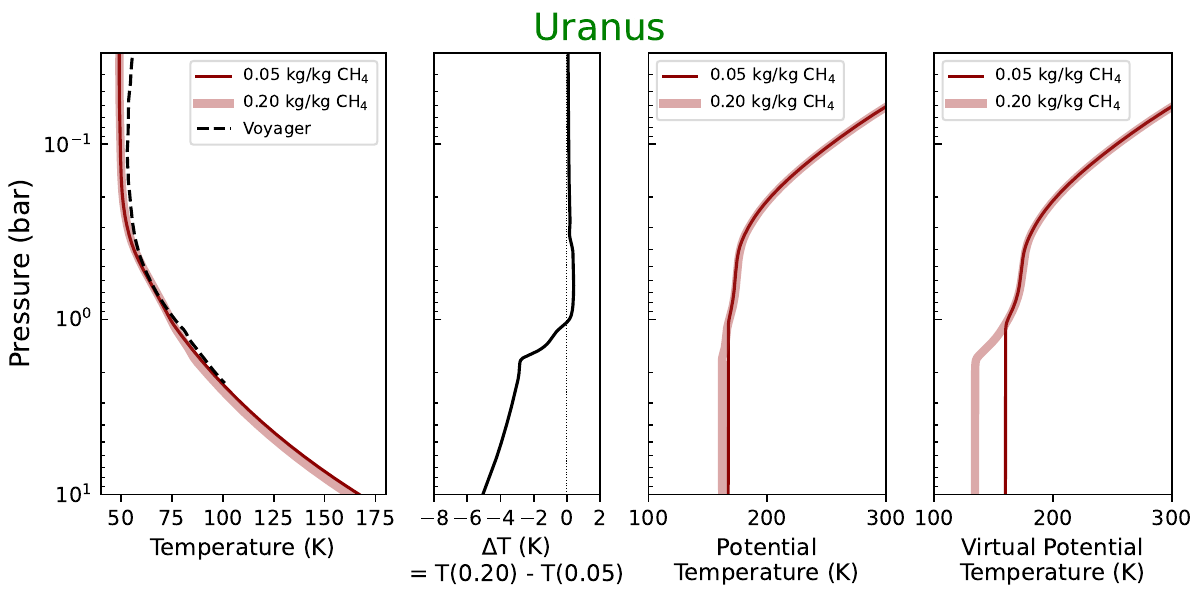}
  \label{fig:u_2_sim}
\end{subfigure}
\begin{subfigure}{0.5\textwidth}
  \centering
  \includegraphics[width=\linewidth]{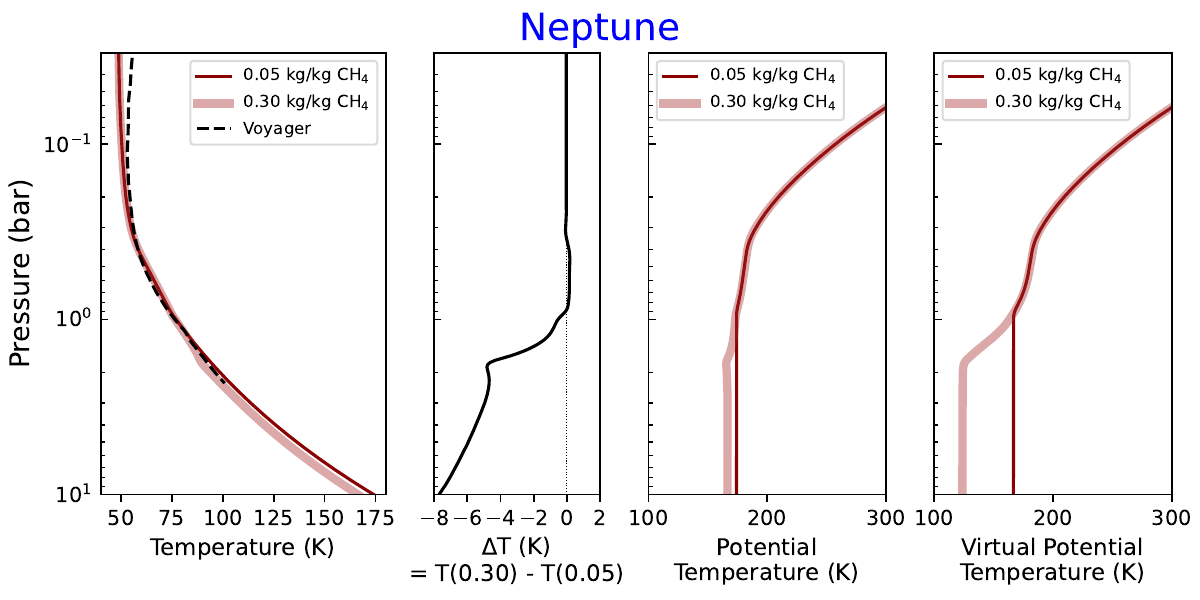}
  \label{fig:n_2_sim}
\end{subfigure}
\caption{Temporally and horizontally averaged temperature profile, temperature difference between two simulations, potential and virtual potential temperature profile from simulations.}
\label{fig:u_and_n_2_sim}
\end{figure*}

\begin{figure*}[ht!]
    \centering
    \includegraphics[width=\linewidth]{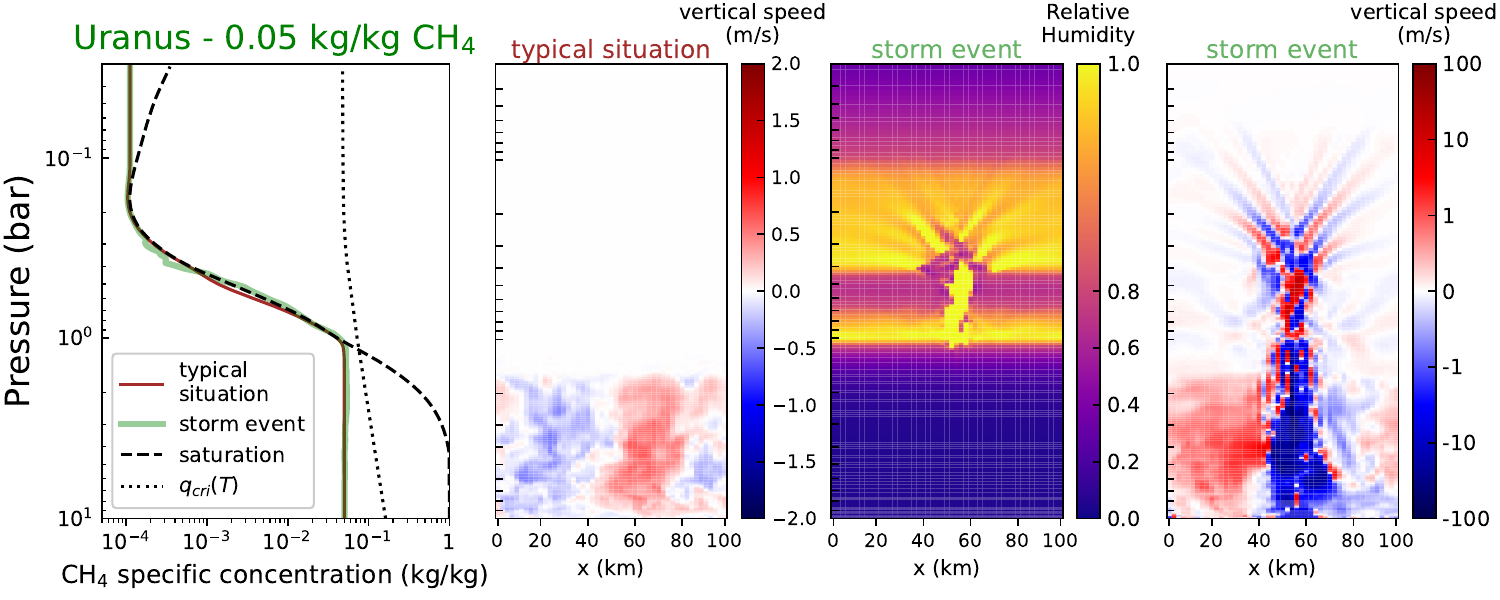}
    \includegraphics[width=\linewidth]{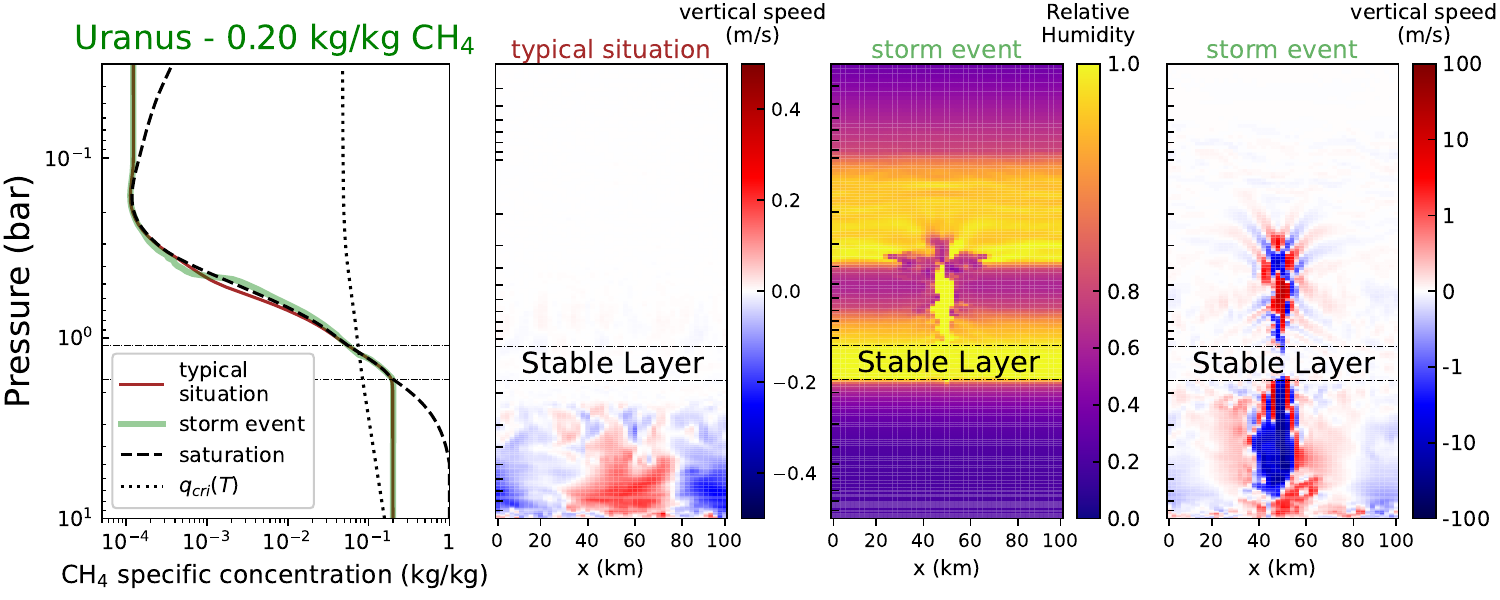}
    \caption{Uranus - Simulations synthesis.
    Simulations with 0.05~kg/kg methane ($\sim$ 0.8\% methane in number) and 0.20~kg/kg methane ($\sim$ 3.6\% methane in number) in the deep atmosphere are presented.
    The first plot shows methane specific concentration averaged over the domain for the "typical situation" profile and taken at a given column of the model for the "storm event" profile.
    \\
    For the speed colormap of the "typical situation", we have chosen an arbitrary snapshot that illustrates well what the "typical situation" looks like. This "typical situation" is sometimes perturbed, a convective storm occurs and the atmosphere has the "storm event" structure. For the colormap of the "storm event" structure, we have chosen an arbitrary snapshot of a convective storm among the ones occurring during the whole simulation.
    \\
    On the fourth panel, the color bar for vertical speed is linear in the range [-1,1] and becomes logarithmic in the ranges [-100,-1] and [1,100].}
    \label{fig:Uranus_simulations}
\end{figure*}

\begin{figure*}[ht!]
    \centering
    \includegraphics[width=\linewidth]{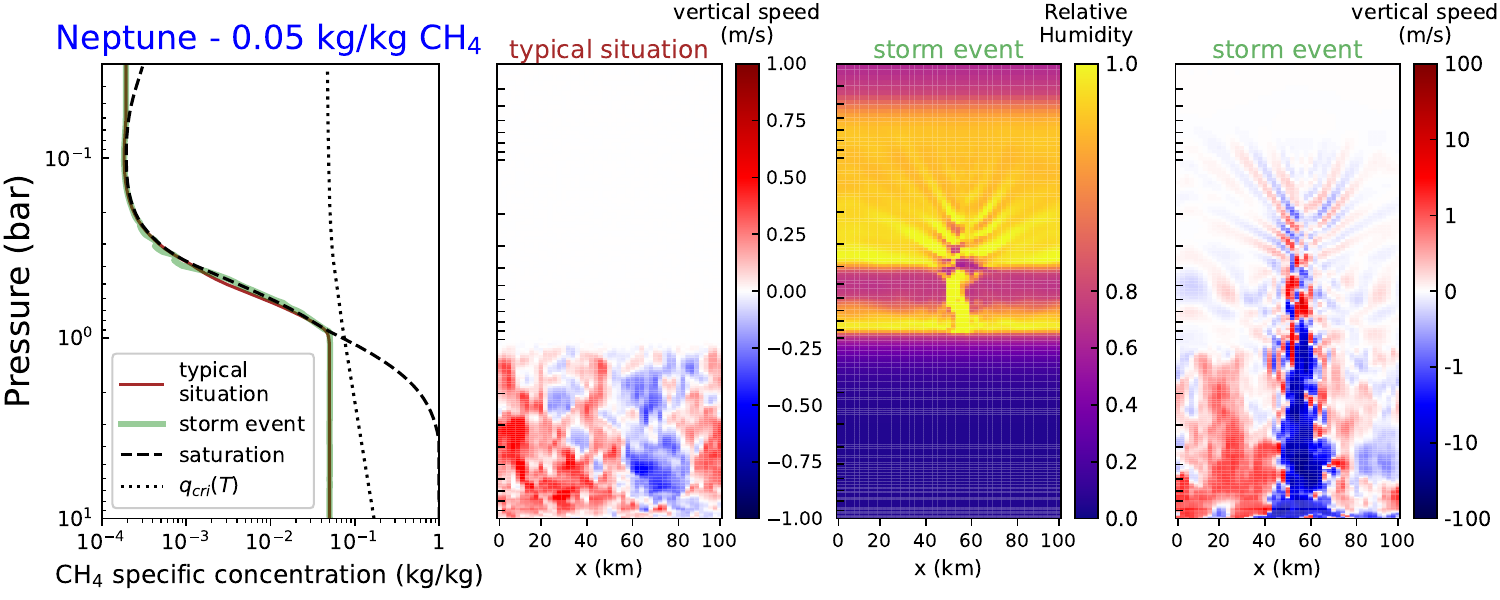}
    \includegraphics[width=\linewidth]{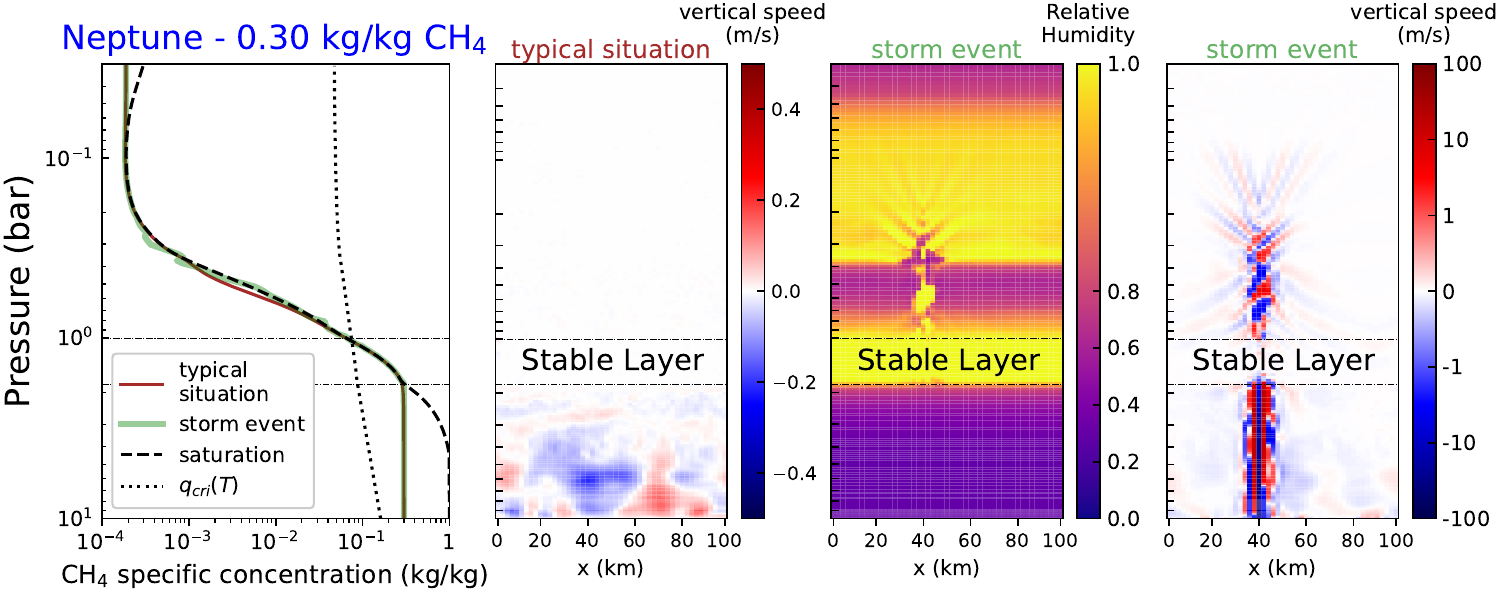}
    \caption{Neptune - Simulations synthesis.
    Simulations with 0.05~kg/kg methane ($\sim$ 0.8\% methane in number) and 0.30~kg/kg methane ($\sim$ 6.2\% methane in number) in the deep atmosphere are presented.
    Same layout as in Figure \ref{fig:Uranus_simulations};}
    \label{fig:Neptune_simulations}
\end{figure*}

In this section, we describe the thermal structures arising in our simulations as well as dynamics and convective activity, then we highlight the role of convection inhibition in those properties.

\subsection{Mean temperature profiles, dynamics and convective activity}

Temperature profiles allow us to check if the simulations are capable of capturing the phenomena we want to study, and if the model provides results in line with observations.
In this section, we look at the temporally and horizontally averaged temperature profiles of the 3D simulations.

Mean temperature profiles are close to \textit{Voyager 2} profiles, and similar to those obtained by \citet{Milcareck2024}.
Above condensation levels, temperature profiles are moist-adiabatic.
In simulations where we expect inhibition of moist convection and the formation of a stable layer (Uranus 0.20 kg/kg CH$_4$ and Neptune 0.30 kg/kg CH$_4$), temperature profiles are super-moist-adiabatic at condensation levels between the 1 and 2 bar levels. However, they remain sub-dry-adiabatic (see Figure \ref{fig:u_and_n_2_sim}).
This structure is intimately related to the initialization profiles due to the long radiative timescale.
Another recent study of the tropospheres of ice giants by \citet{Ge2024} exhibits a super-dry-adiabatic profile at these levels. We come back to these differences in Section \ref{discussion}.
Profiles in \citet{Ge2024} have similar temperature ranges for Neptune, although they are a few Kelvin colder.

\par
The potential temperature of a fluid parcel at a given pressure is the temperature the parcel would reach if it were brought adiabatically to a standard reference pressure (the 10-bar pressure at the bottom of the model in our case). In this study, the virtual potential temperature is more useful.
Virtual potential temperature $\theta_{\text{v}}$ is the theoretical potential temperature $\theta$ of dry air that would have the same density as moist air:

\begin{equation}
    \theta_{\text{v}} = (1- (1-\frac{1}{\epsilon_{\text{CH}_4}})q_{\text{v}})~\theta
\end{equation}

\noindent
where $\epsilon_{\text{CH}_4}=\frac{M_{\text{CH}_4}}{M_\text{gas}}=\frac{16.04}{2.3}=6.97$ is the ratio between the molecular weight of the condensable species, in this case methane, and the molecular weight of non-condensing air; $q_{\text{v}}$ (kg/kg) is the vapor specific concentration.
As $\epsilon_{\text{CH}_4}$ is greater than 1, $\theta_{\text{v}}$ is lower than $\theta$.
\par
Variations of $q_{\text{deep}}$ from one simulation to another show no difference on the average temperature profiles above the 1 bar level (see Figure \ref{fig:u_and_n_2_sim}).
Below the 1 bar level, simulations with a high $q_{\text{deep}}$ are colder than simulations with a low $q_{\text{deep}}$. These average temperature profiles are highly dependent on the initial 1D profile. The absorption and emission of methane, and its impact on radiative-convective equilibrium, may explain these differences in temperature gradients.
\citet{Ge2024} have also shown that methane enrichment in the deep atmosphere leads to colder temperature profiles.
Virtual potential temperature profiles also show differences, in line with expectations.
For simulations with $q_{\text{deep}}=$ 0.20 and 0.30~kg/kg, condensation level is reached deeper than for simulations $q_{\text{deep}}=0.05$ kg/kg.
We expect dry convection to extend higher in simulations with $q_{\text{deep}}=$ 0.05~kg/kg. All simulations share the same global 3-layer structure (see Figure \ref{fig:Uranus_simulations} for Uranus and Figure \ref{fig:Neptune_simulations} for Neptune):
\begin{itemize}
    \item a dry layer between 10 and 1-2 bars, where the virtual potential temperature is almost constant (Figure \ref{fig:u_and_n_2_sim}). In this layer methane abundance is almost constant, and equals the value we have fixed in the deep atmosphere. Regions with constant virtual potential temperature correspond to well-mixed layers (CH$_4$ remains at the same concentration) close to the dry-adiabatic gradient. Dry convection permanently occurs in this layer, with typical speeds of about 1 m/s, as shown by the "typical situation" plot (second panel of Figures
    \ref{fig:Uranus_simulations}
    and \ref{fig:Neptune_simulations}). Downdrafts and updrafts can be observed. This is a mixing layer. The upper part of this layer may be non-convective as a thin methane gradient is formed to link the two constraints that delimit this layer. The constraint at the bottom is the $q_{\text{deep}}$ quantity at 10 bars which is transported upward by convection, and the constraint at the top is the condensation level.
    \item a moist layer between 1-2 bars and 0.1 bar, where methane abundance is close to the saturation vapor curve and virtual potential temperature slightly increases. This layer is the moist troposphere where moist convection episodically occurs. Most of the time, no convection occurs in this layer, as shown by "typical situation" plots. Sometimes, convective storms occur, as shown by "storm event" plots (fourth panel of Figures
    \ref{fig:Uranus_simulations}
    and \ref{fig:Neptune_simulations}).
    Convective storms are characterized by positive speeds in the moist layer (that can be higher than 10 m/s), and a strong downdraft in the dry layer (the large blue cell on the fourth panel of Figures \ref{fig:Uranus_simulations} and \ref{fig:Neptune_simulations}). Gravity waves can be identified above the storm. They transport energy.
    When a storm occurs, the moist layer is saturated and relative humidity reaches 100\% in most of the vertical levels of the moist layer where the storm is located.
    \item an upper layer above the 0.1 bar level up to the top of the model at 0.03 bar, where virtual potential temperature increases more strongly. No convection is expected in this layer but gravity waves can propagate upward.
\end{itemize}

The simulated atmosphere alternates between a "typical situation" structure and a "storm event" structure.
In the "typical situation" structure, there is an expected local minimum of relative humidity in the moist layer. The temperature profile of this layer is confined between the dry-adiabatic temperature gradient and the moist-adiabatic temperature gradient. The temperature gradient is not steep enough for dry convection and the layer is not saturated for moist convection to occur. This layer remains as such until it reaches saturation and convective storms occur. The condensation level below and the cold trap above surround this minimum of relative humidity.

\subsection{Convection inhibition in simulations}

\begin{figure*}[ht!]
\begin{subfigure}{\textwidth}
    \begin{subfigure}{0.49\textwidth}
        \centering
        \includegraphics[width=\linewidth]{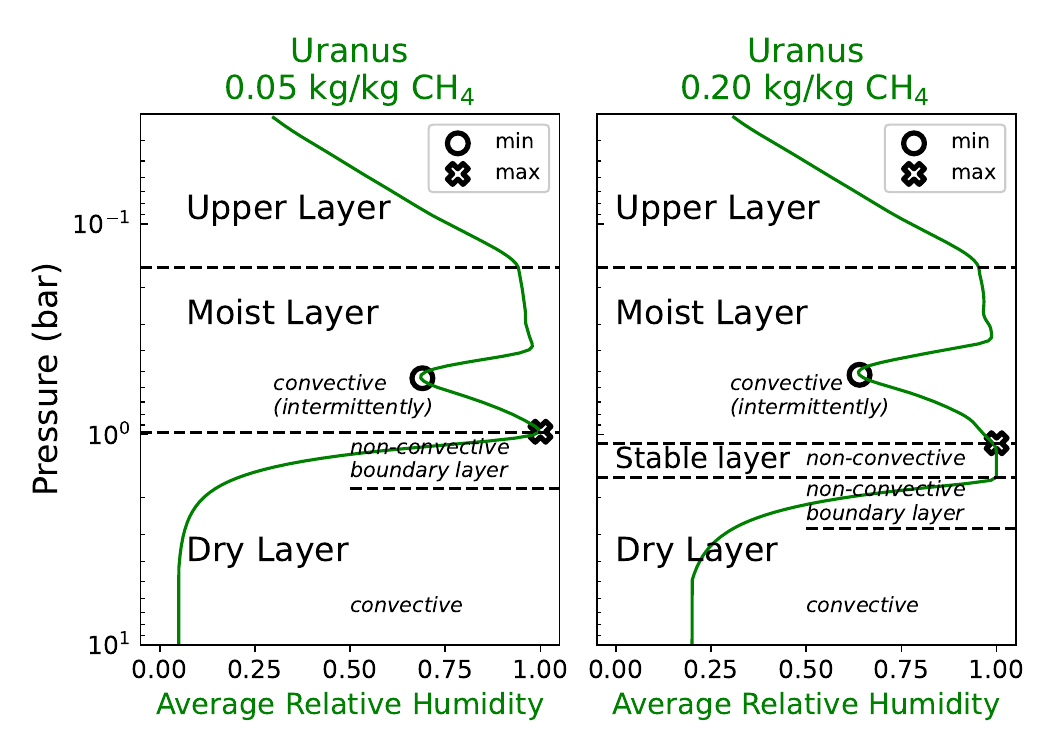}
    \end{subfigure}
    \begin{subfigure}{0.49\textwidth}
        \includegraphics[width=\linewidth]{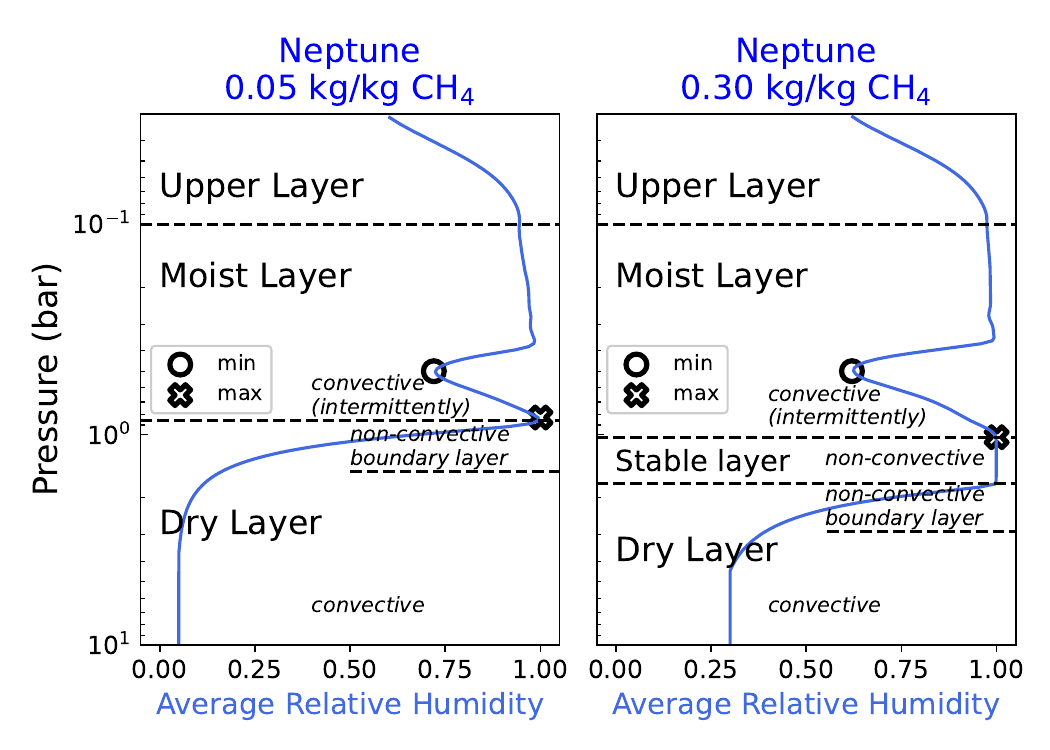}
    \end{subfigure}
\end{subfigure}
\begin{subfigure}{\textwidth}
    \begin{subfigure}{0.49\textwidth}
        \includegraphics[width=\linewidth]{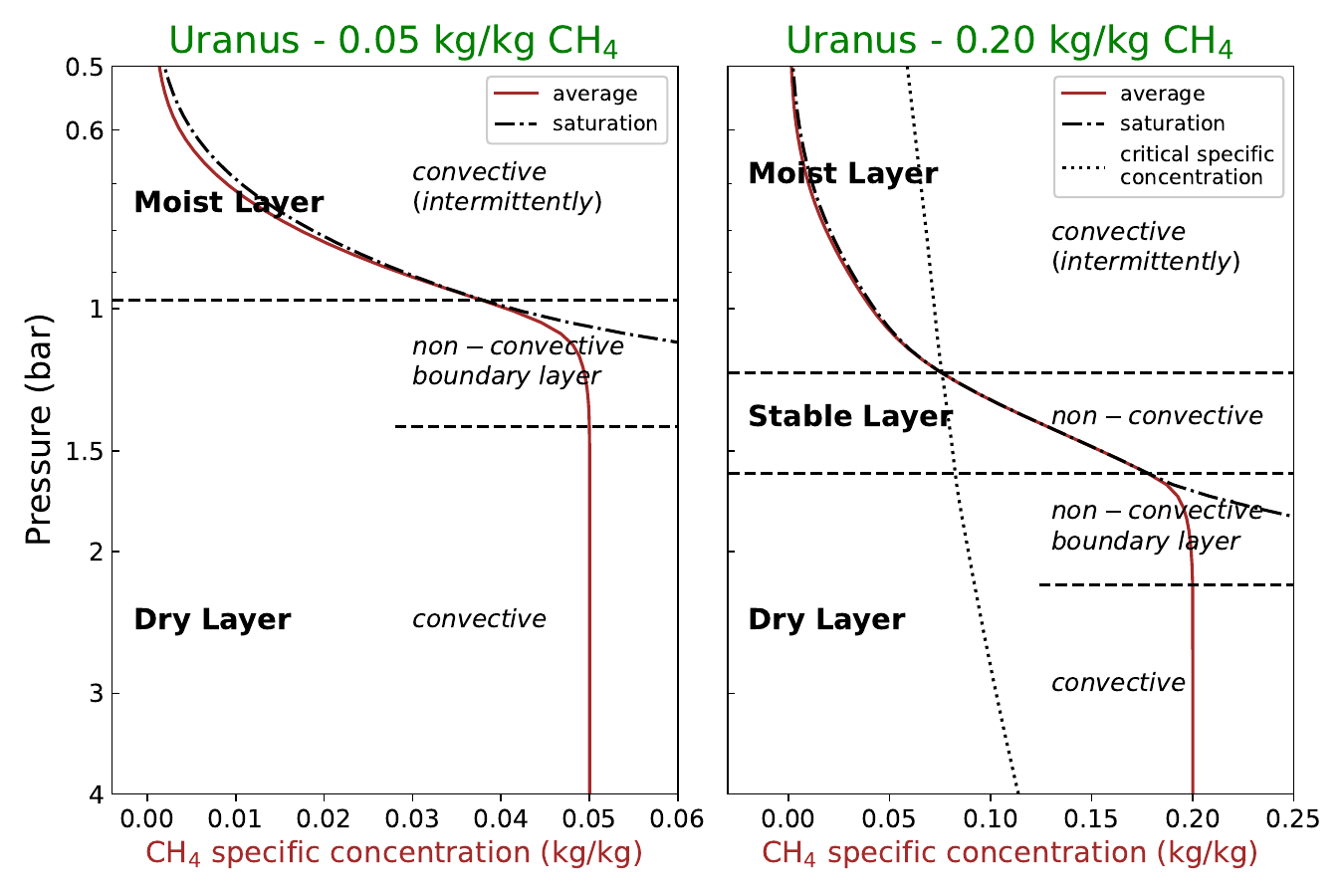}
    \end{subfigure}
    \begin{subfigure}{0.49\textwidth}
        \centering
        \includegraphics[width=\linewidth]{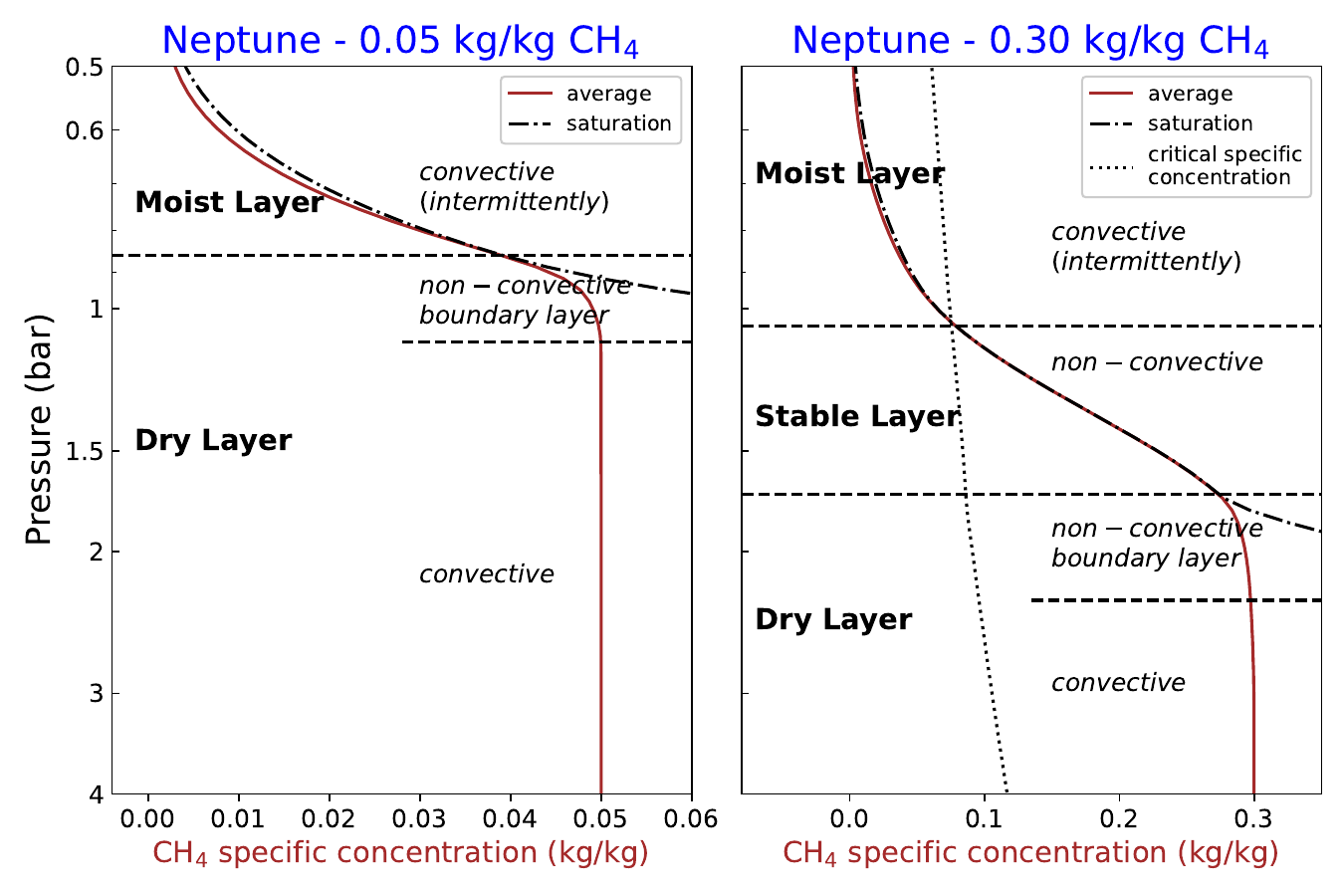}
    \end{subfigure}
\end{subfigure}
\caption{Average structures observed in Uranus and Neptune simulations. \\
First line: Average relative humidity in Uranus and Neptune simulations. The markers (O) and (X) indicate the minimum and maximum relative humidity levels in the moist layer, which will be studied in the next section. \\
Second line: Zoom between 0.5 bar and 4 bars. We plot the methane specific concentration and we choose a linear scale (instead of log-scale as before). \\
A 3-layer structure appears for all simulations with a "Dry Layer" (dry troposphere), a "Moist Layer" (moist troposphere), and an "Upper Layer" (bottom of the stratosphere). A 4th "Stable Layer" appears when the methane abundance in the deep atmosphere exceeds the critical specific concentration. The analytical criteria explain the convective and non-convective parts of these layers.
\\
The zoom (bottom panels of Figure \ref{fig:3_layers}) highlights the methane gradient that exists in the upper part of the dry layer just below condensation level and that forms a "non-convective boundary layer". On the panels of zoom plots where $q_{\text{deep}}=$ 0.05~kg/kg, we do not see the critical specific concentration because the plotted range is lower than it.}
\label{fig:3_layers}
\end{figure*}

The analysis of the dynamics and convective activity in the 3-layer structure highlights that convection can be inhibited.
In the dry layer, methane abundance is constrained at both ends of the layer:
by the fixed methane abundance $q_{\text{deep}}$ in the deep atmosphere and by the saturation at the condensation level.
In most of the dry layer, methane abundance is constant and is equal to $q_{\text{deep}}$.
At the top of the dry layer, methane abundance starts to decrease before reaching the condensation level.
This thin part of the dry layer with a methane gradient is a non-convective boundary layer where the criterion for dry convection inhibition applies. This can be particularly seen in the bottom panels of Figure \ref{fig:3_layers} which are zooms of the first panel of Figures \ref{fig:Uranus_simulations} and \ref{fig:Neptune_simulations}. Dry convection stops slightly before the condensation level, marking the top of the dry layer.
This gradient at the top of the dry layer is about 0.5 bar thick. It could be more or less thick as shown by the possible methane gradients in Figure \ref{fig:U_N_methane}. This non-convective layer acts as an obstacle for upward methane transport, limiting the rise of methane from the dry layer to the moist layer. This rise is required for convective storms.

\par
The critical specific concentration for moist convection inhibition ($q_{\text{cri}}(T)$) is plotted (dotted line) on the first panels of Figures \ref{fig:Uranus_simulations} and \ref{fig:Neptune_simulations}.
For simulations run with $q_{\text{deep}}=$ 0.05~kg/kg, on both Uranus and Neptune, methane abundance is always below the critical value $q_{\text{cri}}(T)$ at all pressures. Consequently, if we examine the fourth panel of Figures \ref{fig:Uranus_simulations} and \ref{fig:Neptune_simulations} (i.e. "storm event" plots) for 0.05~kg/kg CH$_4$, we can see that moist convection occurs in the entire moist layer. Moist convection is never inhibited.
For simulations run with $q_{\text{deep}}=$ 0.20 and 0.30~kg/kg, the methane abundance may exceed the critical value. A new layer, between the dry layer and the moist layer, appears, with a relative humidity very close to 100\%. This is a non-convective layer (Figure \ref{fig:3_layers}), we call it the "Stable Layer".

In the simulations with $q_{\text{deep}}=$ 0.20 and 0.30~kg/kg, i.e. when the criterion for moist convection inhibition is satisfied, the 3-layer structure becomes a 4-layer structure with this stable moist and non-convective layer appearing between the dry layer and the moist layer.
This non-convective layer is a bit more than 0.5 bar thick in our simulation of Uranus with 0.20~kg/kg methane and almost 1 bar thick in our simulation of Neptune with 0.30~kg/kg methane. The more methane in the deep atmosphere, the thicker this layer is.
It acts both as an obstacle and a reservoir. As moist convection is inhibited in this layer, methane is difficult to transport to higher levels (to levels where moist convection is not inhibited and where convective storms can occur). In addition, the top of this "reservoir" (the level denoted with an X marker in Figure \ref{fig:3_layers}) must be completely saturated in methane to let convective storms occur in the moist layer (see next section).


\section{Formation of convective storms and intermittency} \label{intermittency}

While the previous section focused on averaged structures, we describe here their evolution: how convective storms are formed, what is their frequency and released kinetic energy.

\subsection{Advection of methane: from the dry to the moist layer}

Convective storms are made of rising saturated air.
Figures \ref{fig:Uranus_simulations} and \ref{fig:Neptune_simulations} show the structure of storms. In a few columns of the domain where the storm occurs, the moist layer is filled with methane and reaches 100\% of relative humidity. As the moist layer is confined between the dry-adiabatic gradient and the moist-adiabatic gradient, dry convection can never happen and moist convection happens when saturation is reached.
Convective storms form when enough methane from the deep troposphere is brought to lower pressures, at condensation level where inhibition criteria are not satisfied. When these conditions are gathered, a perturbation can thus create a convective storm.
The mechanical energy of the convective storm is dispersed by gravity waves propagating upward. The convective storm ends with condensed methane precipitating deeper.
\par
Dry convection in the dry layer transports methane to lower pressures. Dry convection faces an obstacle at the top of the dry layer that we call the non-convective boundary layer. To cross this obstacle, methane has to be transported by slow eddy diffusion and has to "climb" the methane gradient. This obstacle slows down the transport of methane to the moist layer and thus controls the frequency of convective storms.
In the simulations with $q_{\text{deep}}$ lower than $q_{\text{cri}}(T)$ (Uranus 0.05~kg/kg CH$_4$ and Neptune 0.05~kg/kg CH$_4$) this gradient below condensation level is slowly becoming smaller with time until it is small enough for a perturbation coming from deeper to reach the moist layer and produce a convective storm.
In the simulations with $q_{\text{deep}}$ exceeding $q_{\text{cri}}(T)$, the stable layer is filling up with methane, acting as a reservoir.
The reservoir becomes full when saturation reaches its top, where methane concentration is lower than its critical value. 
And moist convection can occur.
\par
To quantify methane transport, we estimate the equivalent mixing coefficient (the so-called eddy diffusivity or $K_{zz}$) defined as:
\begin{equation}
K_{zz} \equiv \frac{\left\langle\rho q_\mathrm{v}  w\right\rangle}{\left\langle\rho\partial_z q_\mathrm{v} \right\rangle}
\end{equation}

\noindent
where $\rho$ is the density (kg/m$^3$), $q_\mathrm{v} $ the CH$_4$ vapor specific concentration (kg/kg), $w$ the vertical speed (m/s), $\partial_z$ the partial derivative along the vertical axis, $\left\langle\right\rangle$ the average over horizontal and temporal dimensions.

\begin{figure}[ht!]
    \centering
    \begin{subfigure}{0.23\textwidth}
    \includegraphics[width=\linewidth]{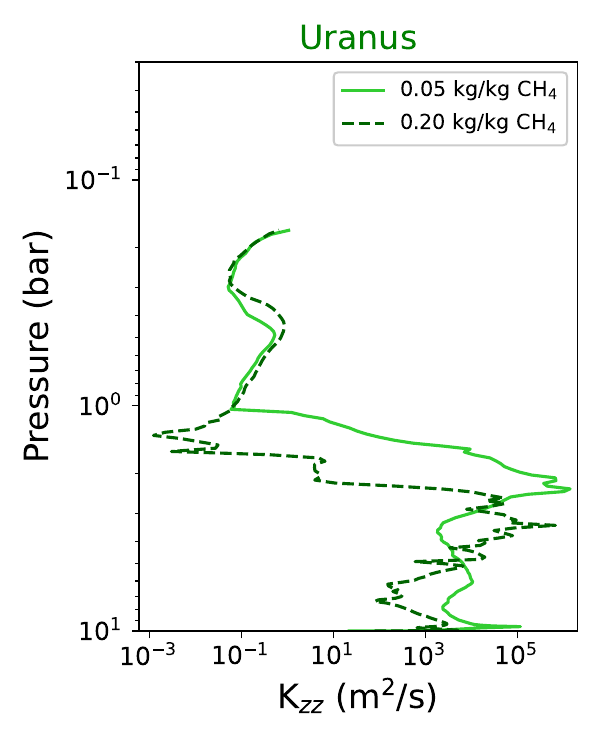}
    \end{subfigure}
    \begin{subfigure}{0.23\textwidth}
    \includegraphics[width=\linewidth]{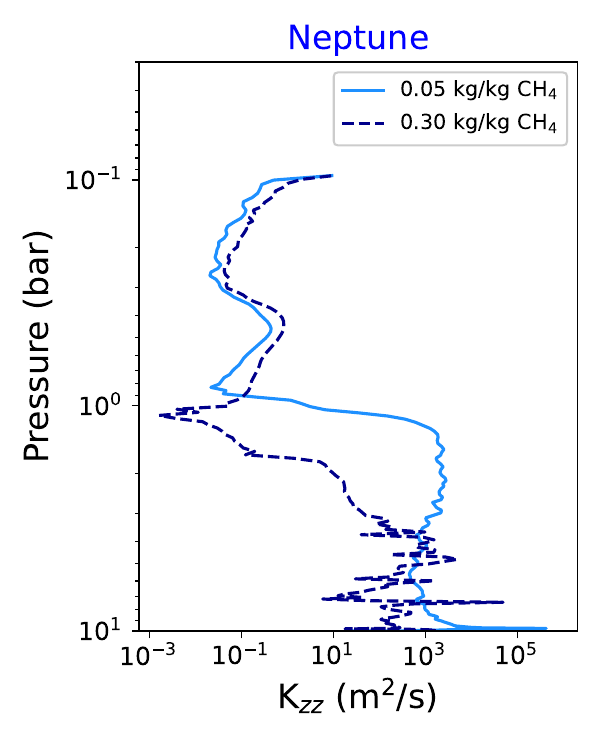}
    \end{subfigure}
    
    \caption{Vertical profile of the equivalent vertical mixing coefficient $K_{zz}$ derived from the simulation}
    \label{fig:kzz}
\end{figure}

The $K_{zz}$ profiles shown in Figure \ref{fig:kzz} are consistent with the dynamics observed in the simulations. In the mixing dry layer methane transport is efficient and $K_{zz}$ is of the order of 10$^2$ to 10$^4$ m$^2$/s. In the moist layer, intermittent convection is illustrated by the peak in the profiles around 0.5 bar and $K_{zz}$ is on the order of 1 m$^2$/s. At the interface between these two layers, there is a low $K_{zz}$ of about 10$^{-1}$ m$^2$/s. $K_{zz}$ is particularly low in the stable layer of the simulations where the criterion for moist convection inhibition is satisfied (Uranus 0.20~kg/kg CH$_4$ and Neptune 0.30~kg/kg CH$_4$).
Using another mesoscale model, \citet{Ge2024} have also calculated $K_{zz}$ profiles that are close to ours: a high $K_{zz}$ in the dry troposphere, a low $K_{zz}$ in the upper layers where convection is inhibited, and a higher $K_{zz}$ in the layer where moist convection can occur.

\begin{figure*}[ht!]
  \centering
  \begin{subfigure}{.24\linewidth}
    \centering
    \includegraphics[width = \linewidth]{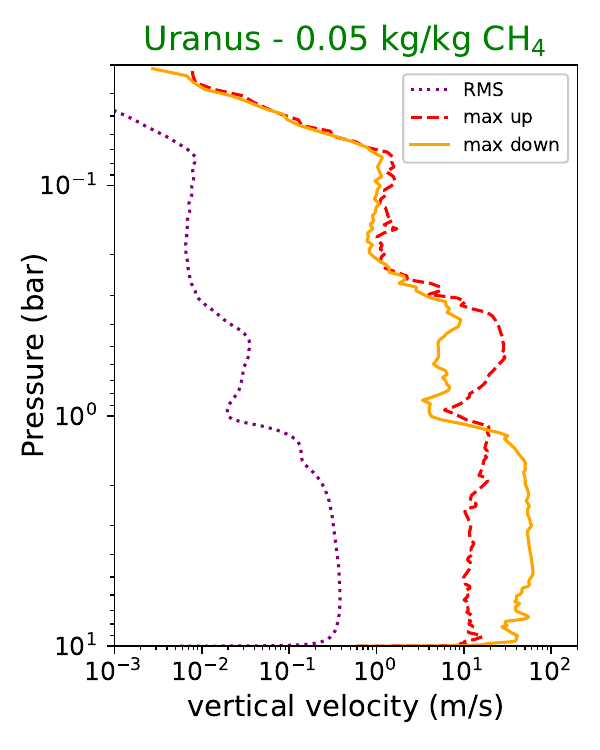}
  \end{subfigure}
  \begin{subfigure}{.24\linewidth}
    \centering
    \includegraphics[width = \linewidth]{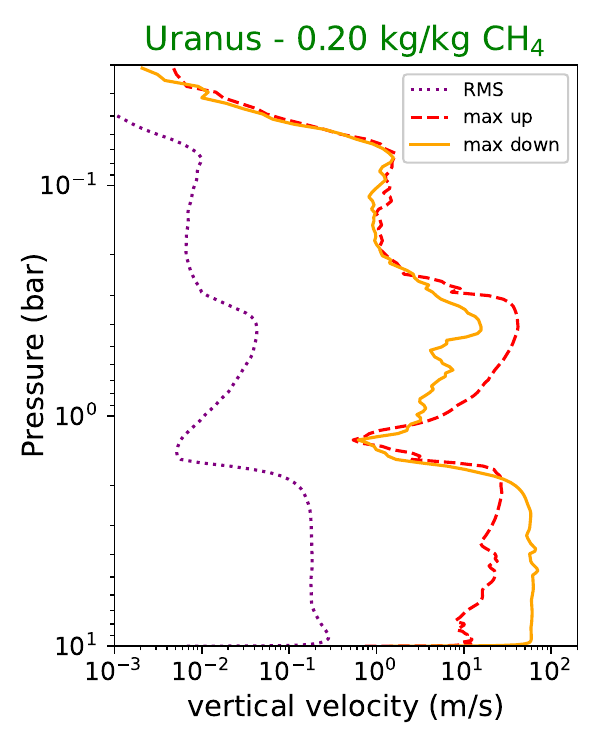}
  \end{subfigure}
  \begin{subfigure}{.24\linewidth}
    \centering
    \includegraphics[width = \linewidth]{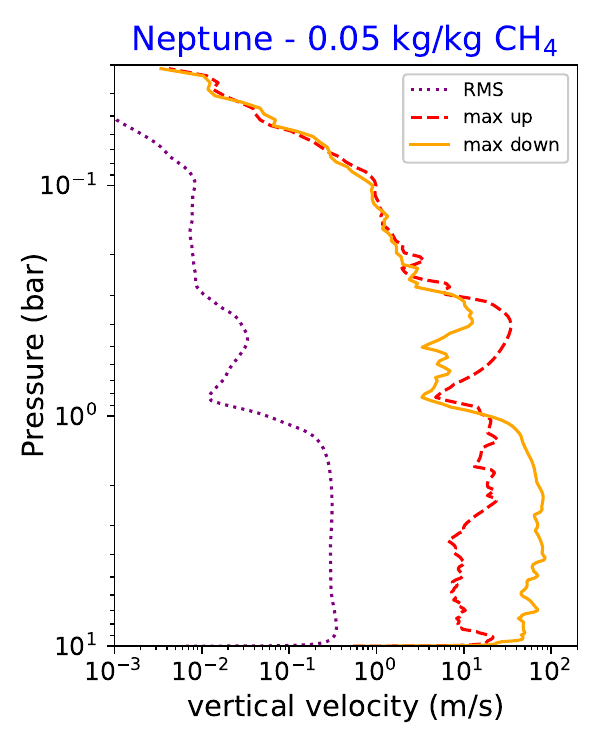}
  \end{subfigure}
  \begin{subfigure}{.24\linewidth}
    \centering
    \includegraphics[width = \linewidth]{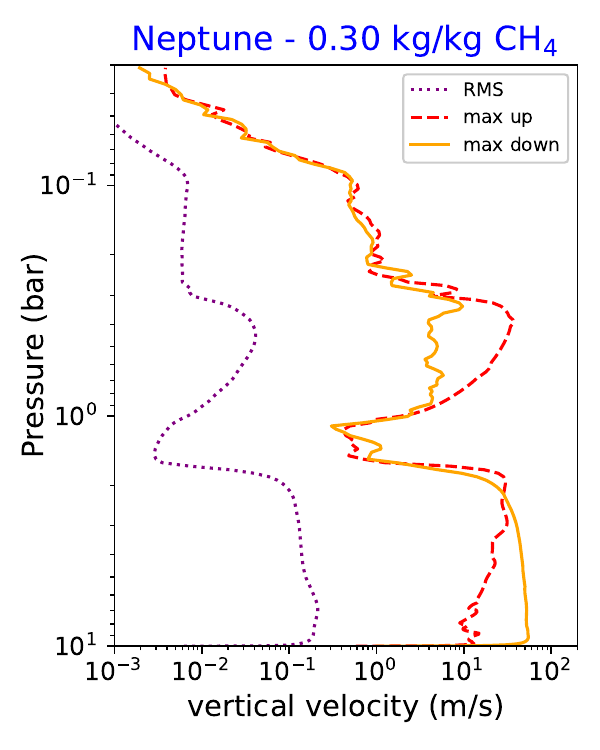}
  \end{subfigure}
  \caption{vertical velocity profiles showing the root-mean-square (RMS, dotted), maximum upward (dashed), and maximum downward (solid) velocities. Averages and maxima values are computed over temporal and horizontal dimensions.}
  \label{fig:vertical_velocities}
\end{figure*}

In the dry layer, below the 2~bar level, maximum downward velocities are higher in absolute value than maximum upward velocities and correspond to the strong downdrafts forming when moist convection occurs (Figure \ref{fig:vertical_velocities}). In the moist layer, maximum upward velocities are higher in absolute value than maximum downward velocities and correspond to the convective storms.
In the simulations run above the critical specific concentration (Uranus 0.20~kg/kg and Neptune 0.30~kg/kg), we can see a strong minimum in the root-min-square (RMS) profiles around 1~bar. This minimum corresponds to the stable layer where the criterion for moist convection inhibition is satisfied. There is no such minimum in simulations below the critical specific concentration. We can see the RMS decreasing slightly before reaching the 1 bar level thus highlighting the non-convective boundary layer at the top of the dry layer, but the local minimum is not very marked.

\subsection{Temporal evolution and intermittence}

The existence of two different structures - a "typical situation" structure and a "storm event" structure - implies intermittency. The previous subsection has explained how the transition between these two structures happens. It works as a cycle.
\par
\begin{figure*}[ht!]
    \centering
    \begin{subfigure}{0.8\textwidth}
    \includegraphics[width=\linewidth]{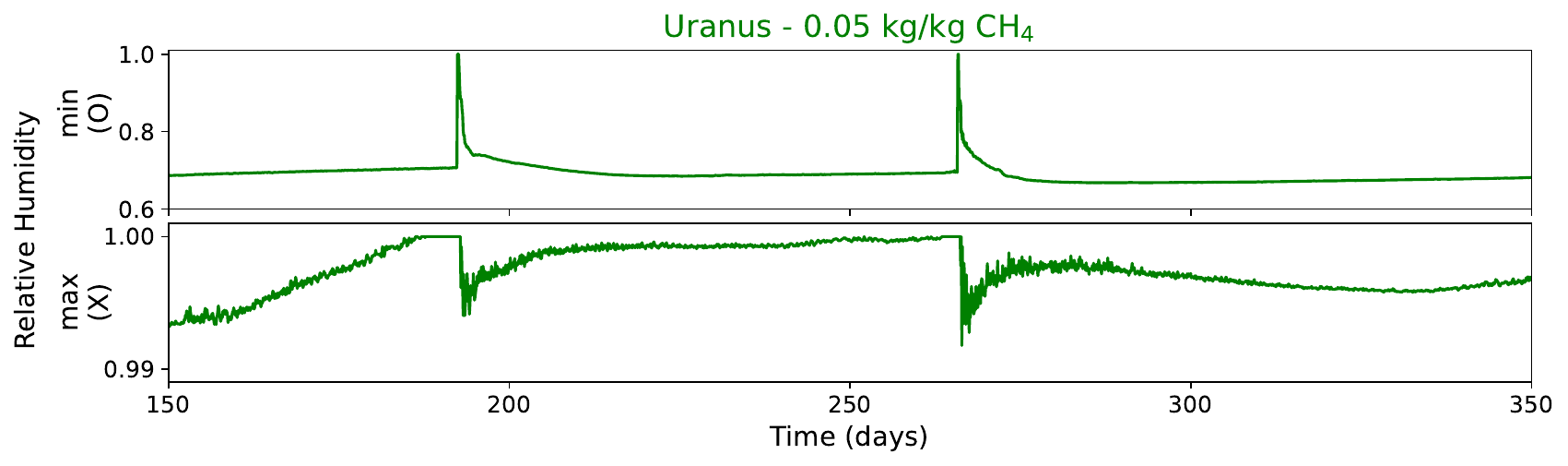}
    \end{subfigure}
    
    \begin{subfigure}{0.8\textwidth}
    \includegraphics[width=\linewidth]{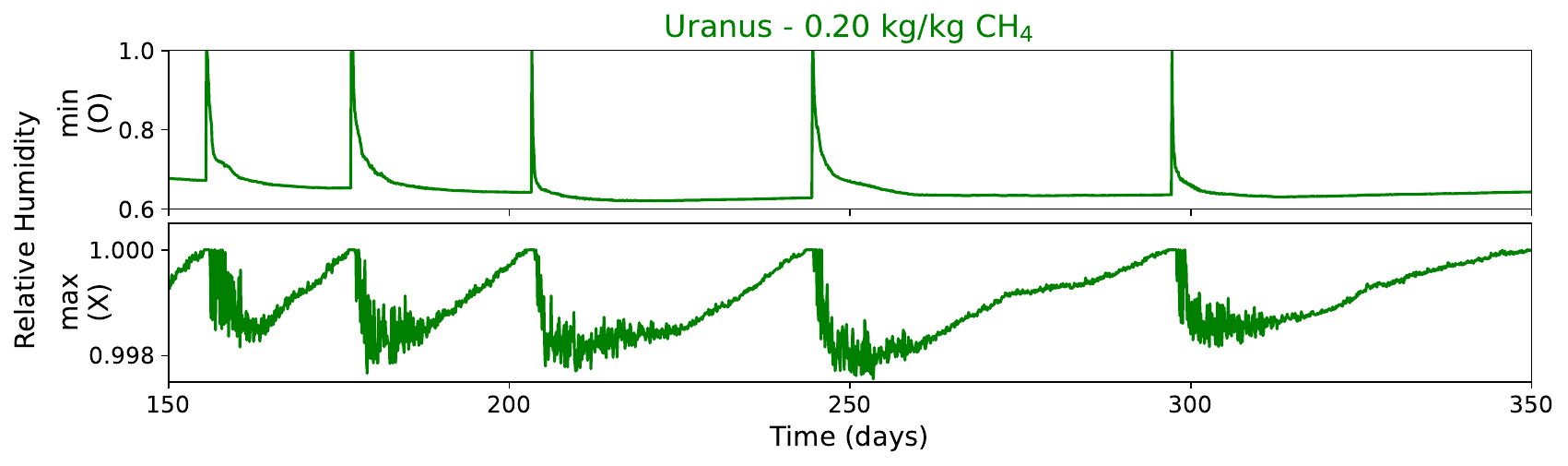}
    \end{subfigure}
    
    \begin{subfigure}{0.8\textwidth}
    \includegraphics[width=\linewidth]{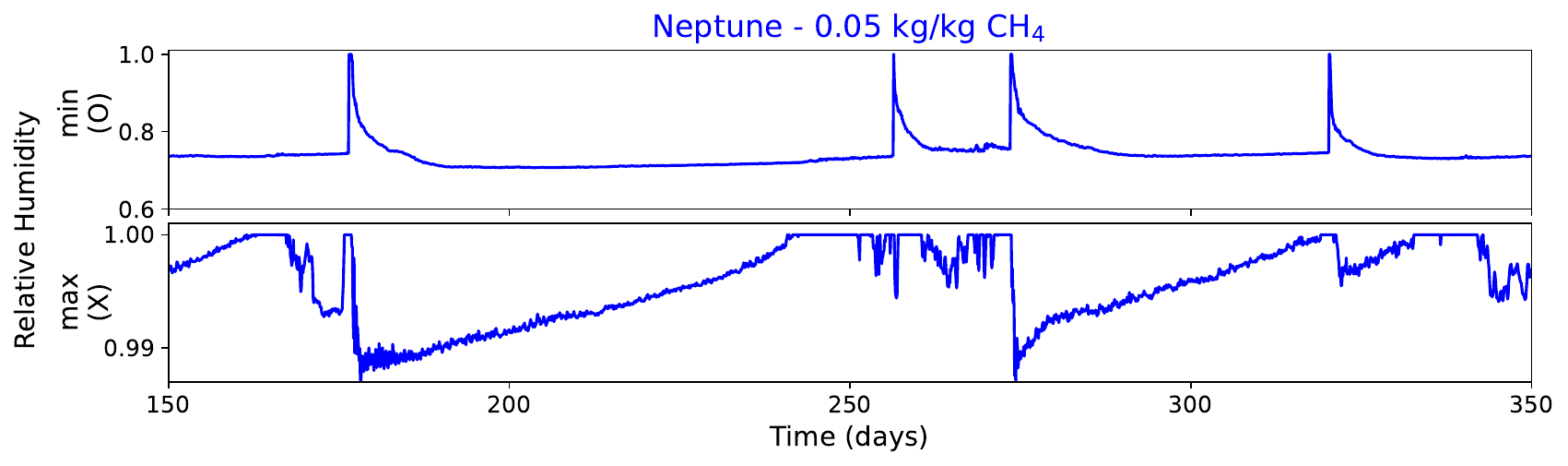}
    \end{subfigure}
    
    \begin{subfigure}{0.8\textwidth}
    \includegraphics[width=\linewidth]{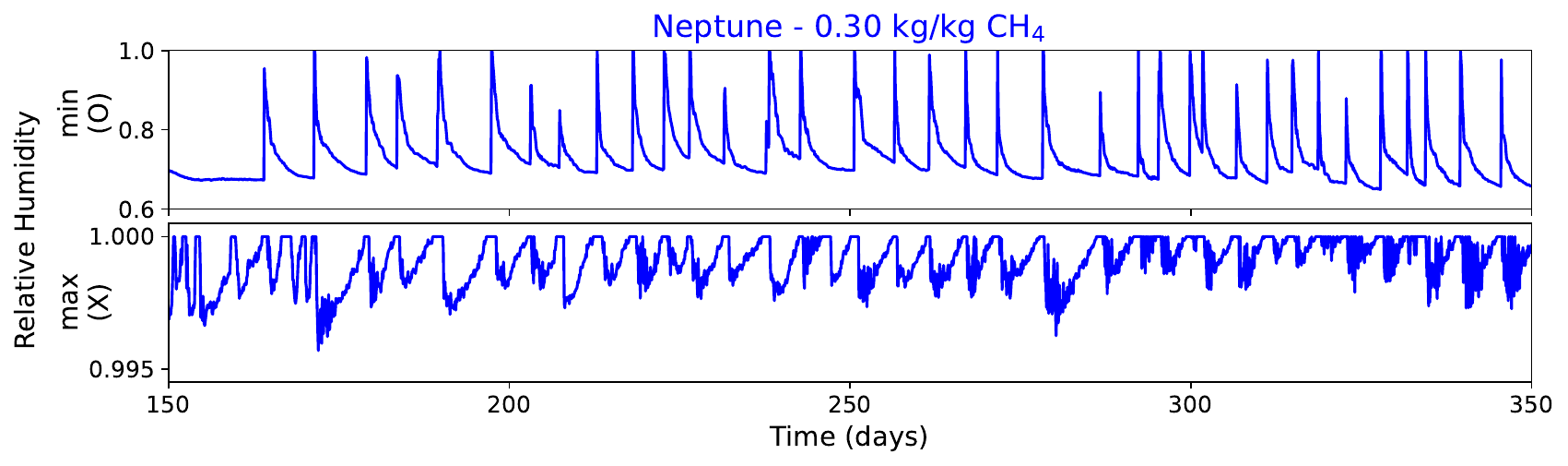}
    \end{subfigure}
    
    \caption{Temporal evolution of the relative humidity in the moist layer at the minimum and maximum levels, respectively identified by the (0) and (X) markers in Figure \ref{fig:3_layers}. For these levels, we keep only the highest value over the domain (latitudinally and longitudinally).}
    \label{fig:uranus_neptune_temporal}
\end{figure*}

To study this cycle, we have identified on the average structures of Figure \ref{fig:3_layers} two important levels in the moist layer: a minimum (indicated by an O marker) and a maximum (indicated by an X marker) of relative humidity. This maximum is also the top of the "reservoir". These levels allow us to identify when a storm occurs. The level of the relative humidity minimum is suddenly filled with methane. In Figure \ref{fig:uranus_neptune_temporal}, we plot the temporal evolution of the relative humidity at these two levels. While minima (O) remain at a constant value for a long time, at the same time maxima (X) are getting filled until they reach 100\%, and then moisture is transferred upward filling the upper levels, among them the minimum level which is more or less the epicenter of the convective storm. In simulations with 0.20 and 0.30~kg/kg CH$_4$, the levels just below the maximum (X) level and that form what we call the "reservoir" are also getting filled. Relative humidity at the maximum level (the top of the reservoir) is always close to 100\% in the simulations exceeding the critical specific concentration, i.e. Uranus 0.20~kg/kg CH$_4$ and Neptune 0.30~kg/kg CH$_4$.

\begin{table*}[ht!]
\centering
\caption{Number of storms and average period between 2 storms in simulations}
\begin{tabular}{|l|c|c|}
     \hline
     \textbf{Simulations} & \textbf{Number of storms in 200 days} & \textbf{Average period between 2 storms} \\
     \hline
     Uranus - 0.05~kg/kg CH$_4$ & 2 & 100 days \\
     \hline
     Uranus - 0.20~kg/kg CH$_4$ & 5 & 40 days \\
     \hline
     Neptune - 0.05~kg/kg CH$_4$ & 4 &  50 days \\
     \hline
     Neptune - 0.30~kg/kg CH$_4$ & 36 & 6 days  \\
     \hline
\end{tabular}
\label{table:storms_calculations}
\end{table*}

The frequency of storms varies a lot from one simulation to another. On the studied window of the last 200 days of the one terrestrial year duration of the simulations, we can count and estimate the period between 2 storms (Table \ref{table:storms_calculations}).
More storms occur in simulations where the critical specific concentration is exceeded: 2.5 times more storms in the simulation of Uranus with 0.20~kg/kg CH$_4$ than in the simulation of Uranus with 0.05~kg/kg CH$_4$, 9 times more storms in the simulation of Neptune with 0.30~kg/kg CH$_4$ than in the simulation of Neptune with 0.05~kg/kg CH$_4$. Storm occurrence is cyclic, with a chaotic aspect (i.e. the cycle is sometimes irregular), which is well illustrated in Neptune's simulation with 0.05~kg/kg CH$_4$ (see the third panel of Figure \ref{fig:uranus_neptune_temporal}), where the time between two storms varies a lot.
Storms are triggered when a sufficiently strong perturbation comes from deeper levels for the moisture of the "reservoir" to be transferred from the levels where moist convection is inhibited to the levels just above where moist convection is no longer inhibited.

\par
Simulations show a much lower frequency of convective storms on Uranus than on Neptune. We attribute this to the absence of internal heat flow. This internal heat flow warms the deep atmosphere in simulations. It is a source of heat at the bottom of the model. There is more energy flux to evacuate from deep layers in Neptune than in Uranus.
One might think that an interesting test would be to turn off the internal heat flow in the settings of Neptune's simulations or to add one in Uranus' simulations. We have done this, but it changes so much the thermal balance so that the temperature gradients, and the conclusions, become unrealistic.
Model and observations agree that Neptune is much more active in producing new cloud systems and cloud variations than Uranus \citep{Hueso2020}.

\subsection{Intensity of convective storms}

\begin{figure*}[ht!]
    \centering
    \begin{subfigure}{0.8\textwidth}
    \includegraphics[width=\linewidth]{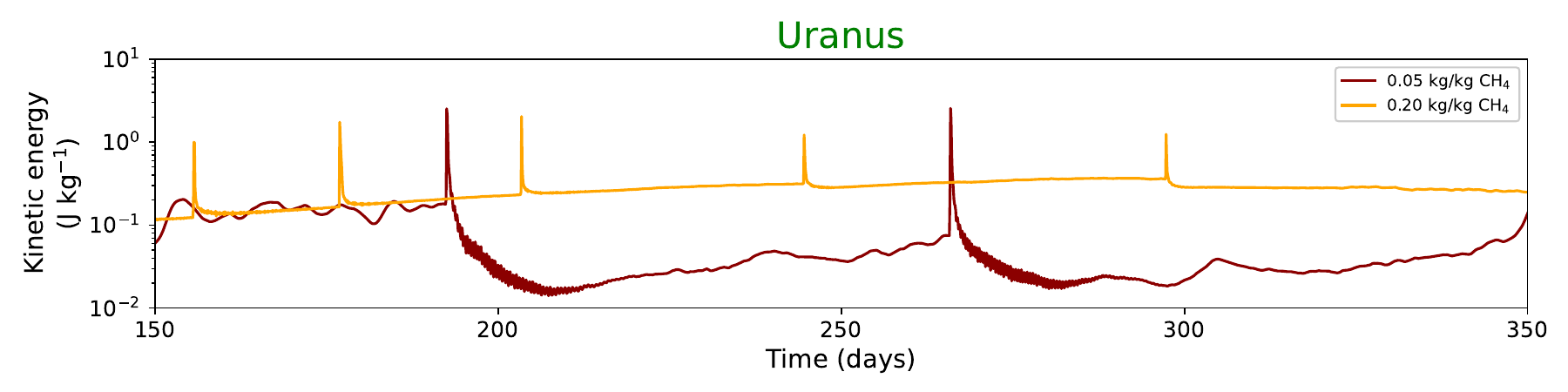}
    \end{subfigure}
    
    \begin{subfigure}{0.8\textwidth}
    \includegraphics[width=\linewidth]{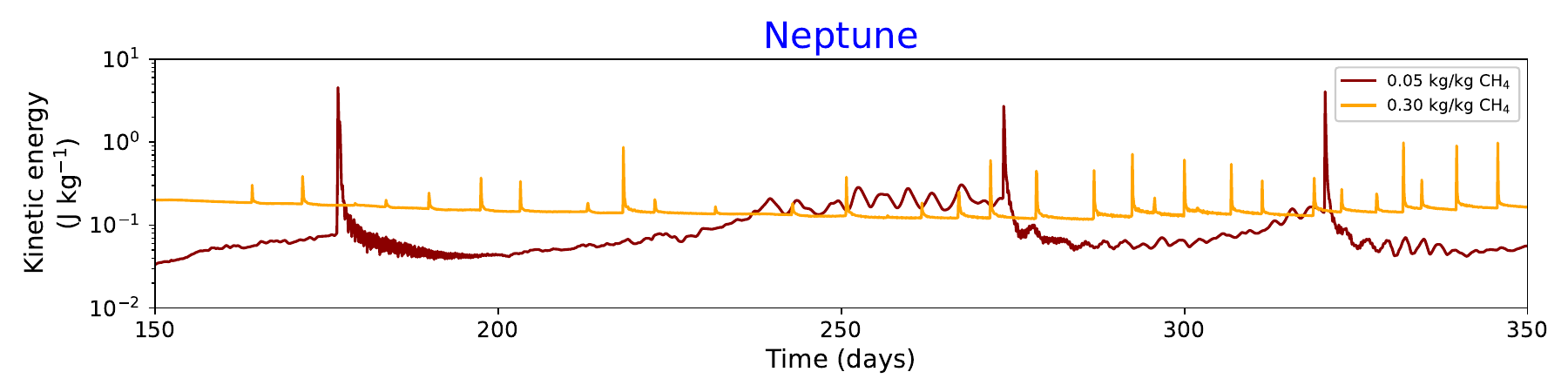}
    \end{subfigure}
    \caption{Kinetic energy variations in the simulations calculated over the whole domain}
    \label{fig:uranus_neptune_temporal_kinetic_energy}
\end{figure*}

The evolution of relative humidity on plots from Figure \ref{fig:uranus_neptune_temporal} has allowed us to identify each convective storm and monitor their frequency. However, these plots do not provide information about the intensity of storms. 
We compute the kinetic energy $E_\mathrm{c}$ per unit of mass on the whole domain of our simulations as follows:

\begin{equation}
    E_\mathrm{c} \text{(J/kg)} = \frac{1}{2}\sum_{i}(u_i^2+v_i^2+w_i^2)\frac{\mathrm{d}m_i}{m_\text{tot}}
\end{equation}

\noindent
where $u_i$, $v_i$, $w_i$ are the three components of velocity of the $i$-th cell of the domain, $\mathrm{d}m_i$ is the mass of this cell, $m_\text{tot}=\sum_{i}\mathrm{d}m_i$ is the total mass of the domain. Figure \ref{fig:uranus_neptune_temporal_kinetic_energy} illustrates the temporal evolution of $E_\mathrm{c}$ in our simulations.

The average kinetic energy of simulations with more methane, i.e. Uranus 0.20~kg/kg CH$_4$ and Neptune 0.30~kg/kg CH$_4$, is of the same order of magnitude as the average kinetic energy of simulations with less methane - Uranus 0.05~kg/kg CH$_4$ and Neptune 0.05~kg/kg CH$_4$.
But when looking at the convective storms that can be identified by the peaks of kinetic energy, we see that the intensity of these peaks is higher in simulations with 0.05~kg/kg CH$_4$. The non-convective layers of these configurations are thinner, but more importantly, the criterion for moist convection inhibition is never met. The release of energy is thus facilitated.
There is a strong correlation between intermittency and intensity. The rarer the storms are the more intense they are. In Figure \ref{fig:uranus_neptune_temporal_kinetic_energy}, the kinetic energy of a storm corresponds to the difference between the peak associated with this storm and the "background" kinetic energy. This "background" kinetic energy is mainly due to dry convection in the troposphere and is of the same order of magnitude from one simulation to another.


\section{Open questions, limitations and suggested scenario} \label{discussion}

\begin{figure}[ht!]
    \centering
    \begin{subfigure}{0.5\textwidth}
    \includegraphics[width=\linewidth]{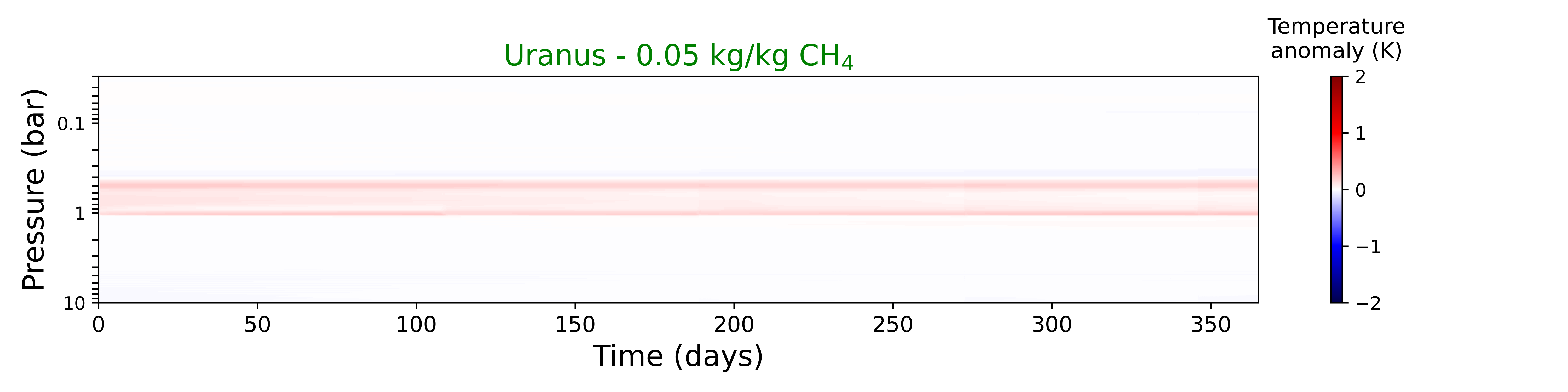}
    \end{subfigure}
    
    \begin{subfigure}{0.5\textwidth}
    \includegraphics[width=\linewidth]{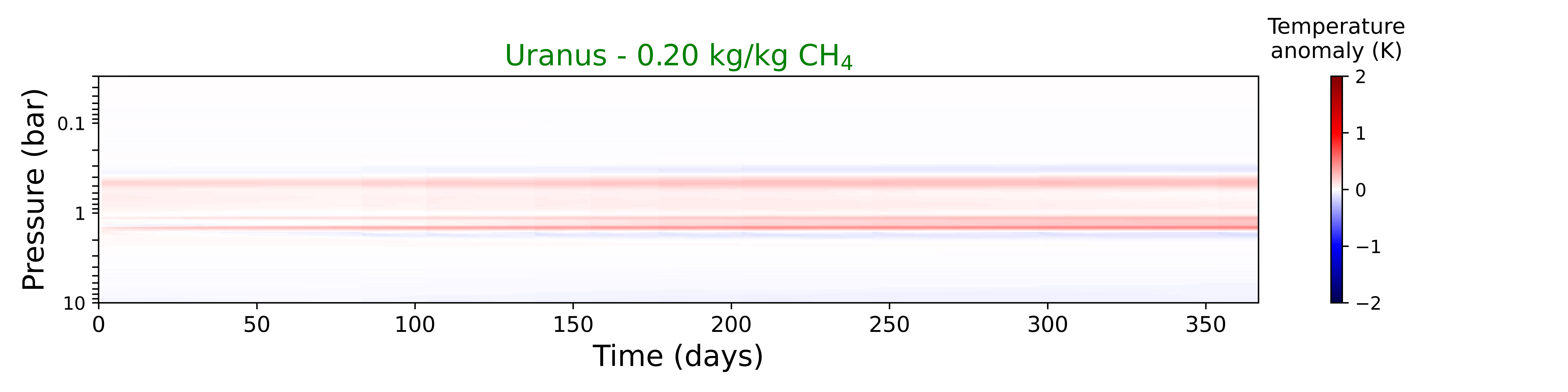}
    \end{subfigure}

    \begin{subfigure}{0.5\textwidth}
    \includegraphics[width=\linewidth]{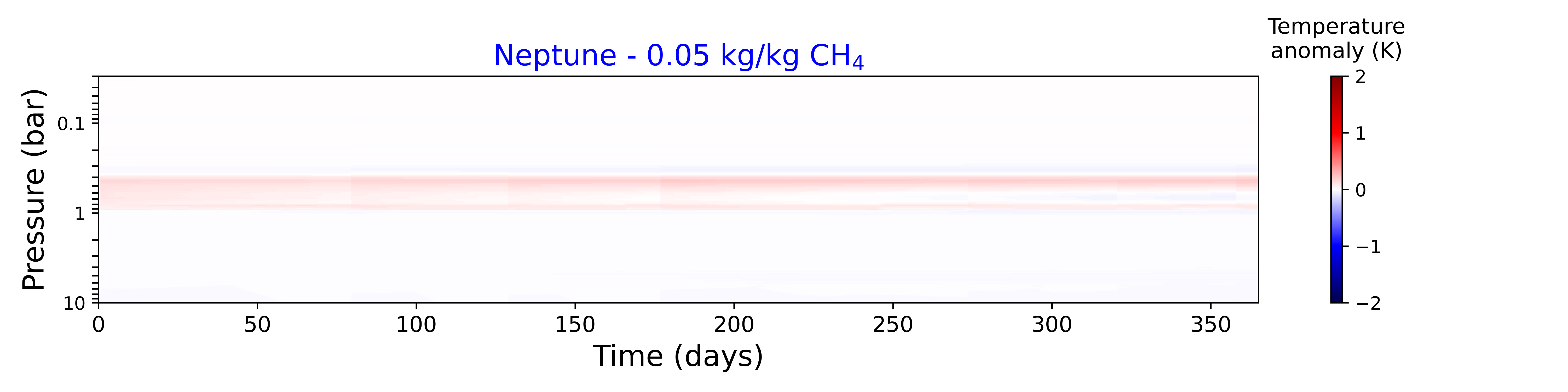}
    \end{subfigure}

    \begin{subfigure}{0.5\textwidth}
    \includegraphics[width=\linewidth]{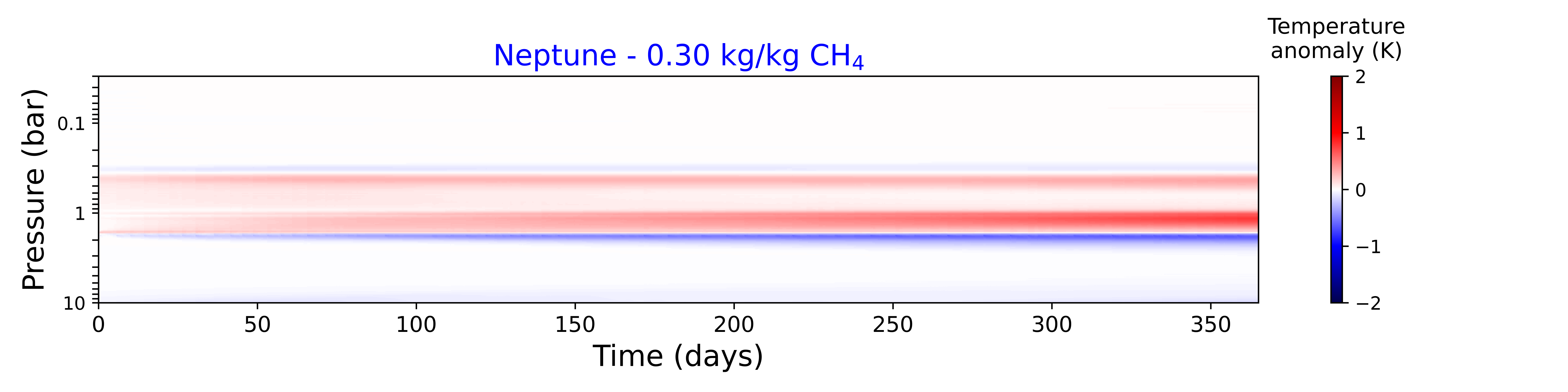}
    \end{subfigure}
    
    \caption{Evolution of temperature anomalies with respect to the initial thermal profile. The temperature anomaly is defined as the difference between the horizontally averaged temperature profile at each time and the initial thermal profile}
    \label{fig:anomalies}
\end{figure}

\begin{figure*}[ht!]
\centering
    \includegraphics[width=\linewidth]{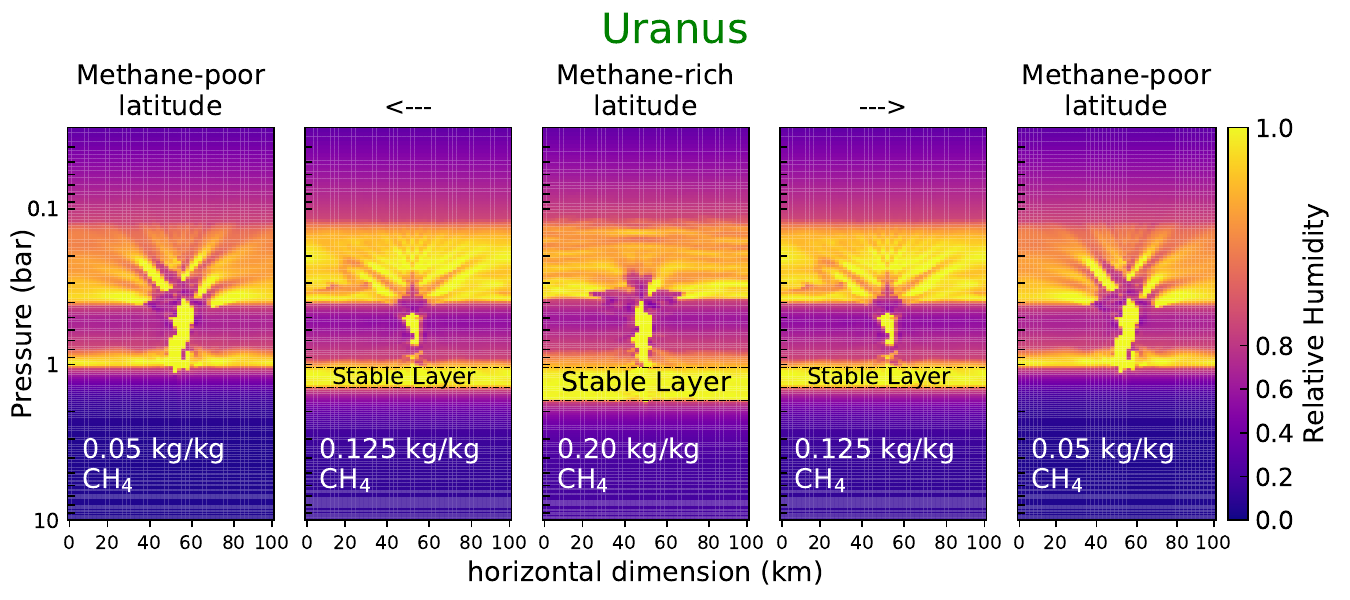}
    \includegraphics[width=\linewidth]{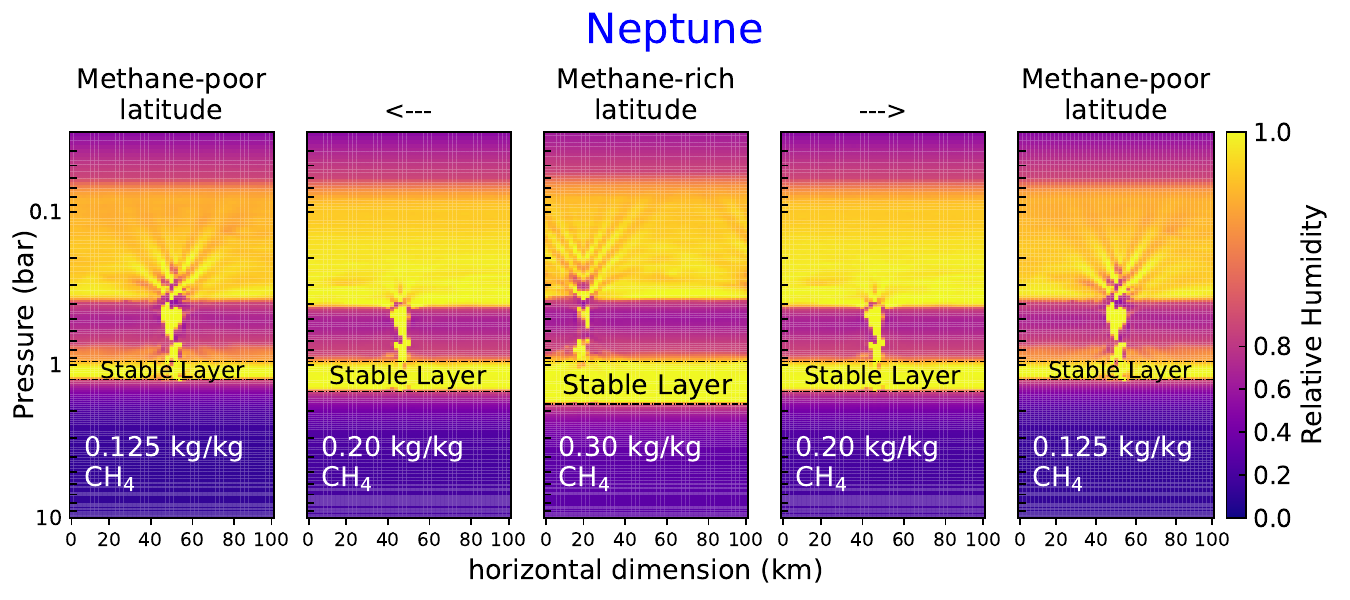}
\caption{Relative humidity (slices of one horizontal dimension) during convective storms, from methane-rich latitudes to methane-poor latitudes, taken from our simulations in ice giants and inspired by latitudinal and vertical variations shown by observations. The simulations are run with different theoretical methane concentrations in the deep atmosphere. \\
Uranus: Observations show latitudinal variations of 1-4\% in methane at 2 bars \citep{Sromovsky2018}, from pole to equator. We plot relative humidity from our simulations with three different methane specific concentrations: 0.05, 0.125, and 0.20~kg/kg. These correspond to volume mixing ratios of 0.8, 2.1, and 3.6\% respectively. \\
Neptune: Observations show latitudinal variations of 2-6\% in methane at 4 bars \citep{Irwin2021}, from pole to equator. We plot relative humidity from our simulations with three different methane specific concentrations: 0.125, 0.20, and 0.30~kg/kg. These correspond to volume mixing ratios of 2.1, 3.6, and 6.2\% respectively. We no longer plot our simulation with $q_{\text{deep}}=$ 0.05~kg/kg since this configuration might not exist on Neptune. \\
We see powerful but rare storms at methane-poor latitudes and weak but frequent storms at methane-rich latitudes. The more methane there is, the more frequent storms are, but also the weaker they are.
\\
To make these plots, we have run 3 more short simulations with $q_{\text{deep}}$ = 0.125 kg/kg CH$_4$ in Uranus, $q_{\text{deep}}$ = 0.125 and 0.20~kg/kg CH$_4$ in Neptune, in addition to our simulations already presented and detailed in this article.
}
\label{fig:conclusion}
\end{figure*}

Here we present the limitations of our model and the associated issues.
Knowing these limitations, we propose a scenario for storm formation in ice giants.\\
\par
\noindent \textbf{Condensates micro-physics}. In our model, the micro-physics of condensation is limited to a very basic scheme. Growth, sedimentation and sublimation of condensates would require a more detailed model. In this first study, we have kept these processes simple and controlled by as few parameters as possible, which at least allows us to test extreme cases that bracket the actual behavior of condensates.
The chosen microphysical parameters limit the potential retention of condensates (e.g. in moist updrafts) to $10^{-10}$ kg of methane ice per kg of air to avoid having to introduce a complex microphysical model. In reality, some condensates can be retained and slow down the updraft. To test whether this would affect our conclusions, we have run simulations where $10^{-3}$ kg/kg of methane ice can be retained before precipitating. As expected, the kinetic energy of moist convective events is reduced, by a factor of 2, when more condensates are allowed, which further favors convection inhibition. At the same time, the frequency of moist convective events is multiplied by a factor 2, demonstrating again that frequency and intensity are strongly correlated.\\

\par
\noindent \textbf{Convergence}.
As discussed in Section \ref{model}, the radiative timescale is too long to enable running the simulation until complete thermal equilibration. As a result, even though we remove the spin-up phase of the simulations (i.e. the first 150 days) and analyze only the part of the simulation where the frequency of storms is rather stable, the thermal structure still evolves slightly over this period. Figure \ref{fig:anomalies} shows the evolution of the temperature anomaly with respect to the initial thermal profile for the 4 simulations. These anomalies remain relatively low and are mostly confined to the stable layer. This is to be expected because the thermal gradient in this region is determined by turbulent diffusion, which is one of the most difficult factors to take into account in the 1D model used for the initialization. The maximum value of the anomaly therefore gives a rough estimate of the uncertainty on the equilibrium temperature profile.\\ 
\noindent
However, we believe that this slight remaining thermal disequilibrium does not affect our main conclusions. First, the layered structure with an inhibition layer has been recovered in a similar setup in \citet{Leconte2024}, even though, in their case, they were able to run their simulation until equilibration. Second, the fact that the size of this inhibition layer increases with deep methane abundance is supported by analytical arguments \citep{Leconte2017}.
A test simulation on Neptune was run after initializing far from saturation (less than 70\% at condensation levels). After 200 days, the stable layer is reformed, demonstrating that its appearance is independent of initialization.
To see whether the evolution of the frequency of storms with abundance was also robust, we carried out a complete set of simulations with a different initial thermal structure: simulations were started from the output of the 1D model where only a dry convective adjustment was performed. These initial conditions are further away from the anticipated equilibrium state of the atmosphere and therefore show larger temperature anomalies during the run. Yet, these simulations show extremely similar behavior in terms of stable-layer sizes and storm frequencies and intensities, which do not differ by more than 50\%.
This confirms that methane turbulent diffusion in the stable layer is the main driver of storm formation.\\

\par
\noindent \textbf{Thermal gradient in stable layers}.
In stable layers, our temperature profiles are super-moist-adiabatic but remain sub-dry-adiabatic. Using less computationally expensive 2D non-hydrostatic simulations that were run for a longer time,
\citet{Ge2024} found a super-dry-adiabatic temperature profile in the stable region and \citet{Leconte2024} found the same structure with a 3D model but for warmer atmospheres
that have shorter radiative timescales. We thus believe that, given more integration time, our thermal profiles would probably converge toward a super-dry-adiabatic one. This however should not affect our conclusions on the occurrence of storms as we have shown that they are driven mainly by the methane cycle and the turbulence in the stable layer, which is only mildly affected by the thermal gradient in the stable layer.\\

\par
\noindent \textbf{Global climate}. Simulations with a general circulation model that accounts for the large-scale dynamics and seasonal changes would provide crucial information.
But we are still facing major unknowns that prevent us from performing fully consistent large-scale simulations: for instance, are the latitudinal variations in methane abundance observed at 1-2~bars on Uranus and 4 bars on Neptune still valid at 10 bars? Is there a pressure at which methane abundance is homogeneous? How do these strong horizontal gradients affect the dynamics? Ideally, small-scale convective simulations should have evolving boundary conditions fed by large-scale simulations while large-scale simulations should include a convective sub-grid scheme derived from small-scale simulations. There is quite some work to be done before reaching such a sophisticated coupling between models. At this early stage, our approach was to develop the small-scale simulations independently and in parallel with large-scale circulation models but to vary the deep methane concentration in order to capture the variety of conditions met at different latitudes. We do believe that by doing so, we were able to capture the mechanisms controlling storms at the scale of our simulations and to provide parametrizations for moist convection inhibition that can be used in global climate models.\\

\par
\noindent \textbf{Diurnal cycle and seasonal variations.}
Our simulations use a constant averaged solar flux and zenith angle and do not include day/night alternation nor the daily and seasonal variations of solar zenith angle, which could have an impact.
The insolation diurnal cycle should have little impact, as radiative time constants are very long: more than 100 terrestrial years at 1 bar.
Concerning seasonal variations of solar flux, the period that we studied (200 days) was small compared to the Uranian and Neptunian years, so the solar flux reaching the planet was almost constant during that time. We could run simulations on the same planet with different constant averaged solar flux corresponding to different latitudes and seasons on the planet. This would change the thermal profile by several Kelvin and the level of methane condensation. It could increase or diminish storm frequency. Because of its obliquity, the case of Uranus is really interesting and the study of seasonal variations will be the subject of future studies.
Our model was run under average insolation conditions, and we tested different values of methane abundance in the deep atmosphere. One could build a pseudo-2D model (i.e. as a function of altitude and latitude) to study latitudinal variations in temperature and methane.\\
\par
\noindent \textbf{Stratospheric methane concentrations.}
Another debated question is how to explain the abundance of methane in the stratosphere.
Observations of \citet{Lellouch2015} show a mixing ratio of CH$_4$ higher than its value at the cold trap in Neptune's stratosphere, while methane abundance strongly decreases with pressure in Uranus' stratosphere. Though we do not seek to explain those particularities in this study, we do not see any methane transport by moist convection to the stratosphere. The overshoots that we simulate do not bring methane into the stratosphere either. The 3D Global Climate Models are more suitable for such a study, as those variations could probably also be caused by large-scale dynamics.
Mesoscale simulations that include the effect of the Coriolis force could also bring new pieces of information.\\

\par
Aware of these limitations, we propose a description of the formation and structure of convective storms for given methane abundances in the deep atmosphere. Our simulations highlight a stable moist and non-convective layer in the regions where the critical specific concentration is exceeded. The thickness of this layer, which is located around the 1 bar level (Figure \ref{fig:conclusion}), depends on the methane abundance in the deep atmosphere ($q_{\text{deep}}$). The frequency and intensity of storms depend on the presence or absence of this stable layer.\\
\indent If the abundance at saturated levels is lower than the critical specific concentration, this stable layer does not exist. This situation is encountered in Uranus, at the poles according to observations. It is illustrated in the two plots done for methane-poor regions, on the edges of Figure \ref{fig:conclusion} (Uranus). A very thin line near the 1 bar level (in yellow on those plots) marks the maximum relative humidity and the separation between the dry layer and the moist layer. Convective storms occur when enough methane is brought up from deeper levels and are intense because moist convection is never inhibited. These intense storms reduce the relative humidity in the levels close to condensation by evacuating methane, and it then takes a long time to "reload" them with methane. The frequency of the storms is lower. \\
\indent If this abundance is higher than the critical specific concentration at saturated levels, a stable layer appears (the larger $q_{\text{deep}}$ the thicker this stable layer). In Uranus, this situation corresponds to the three center plots of Figure \ref{fig:conclusion} (Uranus), and should occur at mid-latitudes and the equator. In Neptune, this situation is encountered in all plots of Figure \ref{fig:conclusion} (Neptune), with the variation of $q_{\text{deep}}$ inducing a variation in the thickness of the stable layer. In this saturated layer of variable thickness, moist convection is inhibited, thus retaining methane and acting as a reservoir. This reservoir needs to be completely full for moist convective storms to occur above. Because of moist convection inhibition, this reservoir is always almost full and the frequency of convective storms is high (about a few days when $q_{\text{deep}}=$ 0.30~kg/kg). Also because of moist convection inhibition, the intensity of the storms is weak: the moist non-convective layers can never be driven upwards by the storm above them. The bottom of the storm is thus never deeper than this level where moist convection is no longer inhibited. The more methane there is, the more frequent storms are, but also the weaker they are.

\par
We have previously explained that convective storms should be more frequent in Neptune than in Uranus, due to the presence of an internal heat flow in Neptune. This might also be reinforced by the fact that the methane abundance in Neptune exceeds the critical abundance for the inhibition of moist convection at all latitudes, while in Uranus the methane abundance at the poles might be lower than this critical abundance. As there is more methane in Neptune than in Uranus, convective storms should be even more frequent (but weaker).
\par
The cyclic occurrence of storms in our simulations may suggest a link with the study by \citet{Li2015} on the frequency of Saturn’s giant storms. The long timescale (about 60 years) and the resolution of convection that they used in their model are quite different from our study. Temperature variations in Saturn's atmosphere are much larger than in the ice giants (we report very few variations in our simulations), and are the main driver of the cyclic occurrence of storms, whereas in our simulations, it is the evolution of the methane profile that is the driver. The frequencies that we have calculated and presented in Table \ref{table:storms_calculations} should be treated with caution. These frequencies depend very much on the assumptions we make about methane microphysics. A better estimate of the frequencies may be obtained by a parametric study of our microphysical parameters. In the current state of our model, even if the frequency of storms is highly uncertain, we can explain the sporadicity of clouds observed on Uranus \citep{Palotai2022,Pater2015} and on Neptune \citep{Karkoschka2011}.
\par
As for the other condensable species in ice giants (H$_2$O, NH$_3$, H$_2$S), we expect the same behavior as for methane. However, some species are less abundant than methane (e.g., NH$_3$, H$_2$S, \citet{Moses2020}), and should never exceed their associated critical mixing ratios for inhibition of moist convection. The behavior of these species could be similar to the simulations presented in this article with 0.05~kg/kg methane. 
In Uranus and Neptune, water should be very abundant in the deep atmosphere below the 100 bars level \citep{Cavalie2017,Venot2020,Moses2020} and should exceed its associated critical mixing ratio. Even if it is difficult to extrapolate our simulated convective regimes in the range of 0.03-10 bars to such high pressures, we could expect behavior similar to the simulations presented in this article with 0.20 and 0.30~kg/kg.
In Jupiter and Saturn, water abundance may exceed the critical mixing ratio associated with water. At certain latitudes, we would then expect behavior similar to the simulations presented in this article with 0.20 and 0.30~kg/kg.


\section{Conclusions}

Using a 3D cloud-resolving model, we have investigated the impact of the change in mean molecular weight due to methane condensation on the formation and inhibition regimes of convective storms in ice giants. Methane being heavier than the H$_2$/He background, its condensation can indeed inhibit convection and moist convective storms.

Observations show both latitudinal variations -  1 to 4\% in Uranus at 2 bars \citep{Sromovsky2014, Sromovsky2018}, 2 to 6\% in Neptune at 4 bars \citep{Irwin2021} - and vertical variations caused by condensation.
\par
Vertical variations at non-saturated levels (i.e. dry levels) strongly stabilize the atmosphere and a super-dry-adiabatic gradient - which is convective in a mixed atmosphere - can remain stable.
The literature \citep{Guillot1995,Leconte2017} highlights the existence of a critical methane abundance at saturated levels. Whether or not this critical abundance is exceeded can inhibit or activate moist convection.
Depending on the form that the methane gradient takes, saturated levels may or may not be above this critical abundance which is 1.2\% at 80~K. It corresponds approximately to the 1 bar level. Latitudinal variations observed at this level are almost all above this critical abundance, but these levels have to be saturated for the criterion of moist convection inhibition to apply.
\par
After having shown in our simulations that this critical methane abundance indeed rules convective storm inhibition and formation, we conclude that:

\begin{itemize}
    \item typical velocities of dry convection in the deep atmosphere are rather low (of the order of 1 m/s), but sufficient to sustain upward methane transport
    \item moist convection at methane condensation level is strongly inhibited
    \item convective storms can form regardless of methane abundance in the deep atmosphere, but they can only form in saturated layers where the methane abundance does not exceed the critical abundance
    \item the formation of convective storms on Uranus and Neptune should be intermittent and follow a loading/unloading cycle
    \item the intermittency and intensity of storms depends on the methane abundance:
    \\
    \indent - where CH$_4$ exceeds the critical abundance in the deep atmosphere (at the equator and the middle latitudes on Uranus, and all latitudes on Neptune), more frequent but weak storms form.\\
    \indent - where CH$_4$ remains below the critical abundance in the deep atmosphere (possibly at the poles on Uranus), storms are rarer but more powerful.\\
    \item storms should be more frequent on Neptune than on Uranus, because of the internal heat flow of Neptune and because there is more methane in Neptune than in Uranus
    \item methane-rich latitudes at saturated (or near-saturated) levels should act as a barrier allowing little energy to be released in one storm, while methane-poor latitudes should allow much more energy to be released in one storm
\end{itemize}

These conclusions could explain the sporadicity of clouds observed in ice giants.
Further observations with the James Webb Space Telescope, which would track moist convective events over a longer observational period or would provide new constraints on methane abundance, could help to bring new insights to the conclusions proposed in this article.


\section*{Acknowledgements}
The authors acknowledge the support of the French Agence Nationale de la Recherche (ANR), under grant ANR-20-CE49-0009 (project SOUND).

\bibliographystyle{aa.bst}
\bibliography{reference}

\end{document}